\documentclass[lettersize,journal]{IEEEtran}

\usepackage[cmex10]{amsmath} 
\usepackage{amsfonts,stmaryrd,acronym,amssymb,amsthm,dsfont}
\usepackage{algorithm}
\usepackage{algorithmic}
\usepackage{graphicx,paralist,mdframed}
\usepackage{url}
\usepackage{cite}
\usepackage[font=footnotesize, justification=raggedright, singlelinecheck=false]{caption}
\usepackage[font=footnotesize, justification=centering, singlelinecheck=false]{subcaption}
\usepackage{optidef}
\usepackage{bbm}
\usepackage{hyperref}
\usepackage{flushend}
\usepackage{enumerate}
\usepackage{pdflscape}
\usepackage[dvipsnames]{xcolor}
\usepackage{mathrsfs}
\usepackage{lipsum}
\usepackage{diagbox}
\usepackage{soul}
\usepackage{kantlipsum}
\usepackage{ragged2e} 
\usepackage{bm}
\usepackage{cuted}
\usepackage{textcomp}
\usepackage{stfloats}

\usepackage[bbgreekl]{mathbbol}
\usepackage[font=small]{caption}
\DeclareSymbolFontAlphabet{\mathbbm}{bbold}
\DeclareSymbolFontAlphabet{\mathbb}{AMSb}%

\graphicspath{{Graphics/}}
\definecolor{ColorHighlight}{rgb}{1,0,0}

\makeatletter
\def\blfootnote{\xdef\@thefnmark{}\@footnotetext}
\makeatother
\newcommand{\indi}[1]{\ensuremath{\mathds{1}}}

\newcounter{mytempeqcounter}

\allowdisplaybreaks
\allowbreak

\newcommand{\calC}{\mathcal{C}}

\newcommand{\calN}{\mathcal{N}}

\newcommand{\bbP}{\mathbb{P}}

\newcommand{\calS}{\mathcal{S}}

\newcommand{\calX}{\mathcal{X}}

\newcommand{\calY}{\mathcal{Y}}

\newcommand{\kld}{\mathbb{D}}

\DeclareMathOperator*{\argmax}{argmax}

\newcommand{\indic}[1]{\ensuremath{\mathds{1}}}

\renewcommand{\leq}{\leqslant} 
\renewcommand{\geq}{\geqslant} 

\acrodef{ACDIS}[ACDIS]{Adaptive Communication Decision and Information Systems}
\acrodef{AEP}{Asymptotic Equipartition Property}
\acrodef{AoA}{Angle of Arrival}
\acrodef{AWGN}{additive white Gaussian noise}
\acrodef{AVC}[AVC]{Arbitrarily Varying Channel}
\acrodefplural{AVC}{Arbitrarily Varying Channels}
\acrodef{PIR-PNSI}{Private Information Retrieval with Private Noisy Side Information}
\acrodef{BER}{Bit-Error-Rate}
\acrodef{BEC}{Binary Erasure Channel}
\acrodefplural{BEC}{Binary Erasure Channels}
\acrodef{BSC}{Binary Symmetric Channel}
\acrodefplural{BSC}{Binary Symmetric Channels}
\acrodef{BPSK}{Binary Phase-Shift Keying}
\acrodef{BICM}[BICM]{Bit-Interleaved Coded-Modulation}
\acrodef{CDF}[CDF]{Cumulative Distribution Function}
\acrodef{CGF}[CGF]{Cumulant Generating Function}
\acrodef{CLT}[CLT]{Central Limit Theorem}
\acrodef{CSI}[CSI]{channel state information}
\acrodef{DMC}[DMC]{discrete memoryless channel}
\acrodefplural{DMC}{discrete memoryless channels}
\acrodef{DMS}[DMS]{Discrete Memoryless Source}
\acrodef{ERM}[ERM]{Empirical Risk Minimization}
\acrodef{FER}[FER]{Frame Error Rate}
\acrodef{ICA}[ICA]{Independent Component Analysis}
\acrodef{iid}[i.i.d.]{independent and identically distributed}
\acrodef{IoT}[IoT]{Internet of Things}
\acrodef{KKT}[KKT]{Karush-Kuhn-Tucker}
\acrodef{LASSO}[LASSO]{Least Absolute Shrinkage and Selection Operator}
\acrodef{LPD}[LPD]{Low Probability of Detection}
\acrodef{LDPC}[LDPC]{Low-Density Parity-Check}
\acrodef{CSCG}[CSCG]{circularly symmetric complex Gaussian}
\acrodef{LLMS}[LLMS]{Linear Least Mean Square}
\acrodef{LMS}[LMS]{Least Mean Square}
\acrodef{MAC}[MAC]{Multiple-Access Channel}
\acrodef{ADSI}[ADSI]{Action-Dependent State Information}
\acrodef{MGF}[MGF]{Moment Generating Function}
\acrodef{MLC}[MLC]{Multi-Level Coding}
\acrodef{MLE}[MLE]{Maximum Likelihood Estimate}
\acrodef{MIMO}[MIMO]{multiple-input multiple-output}
\acrodef{MISO}{multiple-input single-output}
\acrodef{MSD}[MSD]{Multi-Stage Decoding}
\acrodef{MSE}[MSE]{mean-square error}
\acrodef{MMSE}[MMSE]{minimum mean-square error}
\acrodef{PAC}[PAC]{Probably Approximately Correct}
\acrodef{PCA}[PCA]{Principal Component Analysis}
\acrodef{PDF}[PDF]{Probability Density Function}
\acrodefplural{PDF}{Probability Density Functions}
\acrodef{PMF}[PMF]{Probability Mass Function}
\acrodefplural{PMF}{Probability Mass Functions}
\acrodef{PPM}[PPM]{Pulse Position Modulation}
\acrodef{PSD}{Power Spectral Density}
\acrodef{PSK}{Phase Shift Keying}
\acrodef{QKD}{Quantum Key Distribution}
\acrodef{ROC}{Receiver Operating Characteristic}
\acrodef{CVQKD}{Continuous-Variable \ac{QKD}}
\acrodef{QPSK}{Quadrature Phase-Shift Keying}
\acrodef{RV}{random variable}
\acrodef{SIMO}{single-input multiple-output}
\acrodef{SNR}{signal-to-noise ratio}
\acrodef{SINR}{signal-to-interference-plus-noise ratio}
\acrodef{SVM}[SVM]{Support Vector Machine}
\acrodef{TPCP}{Trace-Preserving Completely-Positive}
\acrodef{wrt}[w.r.t.]{with respect to}
\acrodef{WSS}{Wide Sense Stationary}
\acrodef{RHS}{Right Hand Side}
\acrodef{LHS}{Left Hand Side}
\acrodef{PIR}{Private Information Retrieval}
\acrodef{MDS}{Maximum Distance Separable}
\acrodef{LLN}{Law of Large Numbers}
\acrodef{DFRC}{Dual-Function Radar Communication}
\acrodef{ISAC}{Integrated Sensing and Communication}
\acrodef{RadCom}{Joint Radar and Communicatins}
\acrodef{PLS}[PLS]{physical layer security}
\acrodef{RL}{reinforcement learning}
\acrodef{POCS}{projections onto convex sets}
\acrodef{DQN}{deep Q-learning network}
\acrodef{FIFO}[FIFO]{First-In-First-Out}
\acrodef{DDQN}[DDQN]{double deep Q-network}
\acrodef{MDP}[MDP]{Markov decision process}
\acrodef{SGD}{stochastic gradient descent}
\acrodef{DRL}{deep reinforcement learning}
\acrodef{6G}{sixth-generation}
\acrodef{AN}{artificial noise}
\acrodef{IRS}{intelligent reflecting surface}
\acrodef{NOMA}{non-orthogonal multiple access}
\acrodef{CIPC}{channel inversion power control}
\acrodef{QoS}{quality of service}
\acrodef{PGA}{projected gradient ascent}
\acrodef{PGD}{projected gradient descent}

\begin{document}

\title{Resource Allocation for Positive-Rate Covert Communications Using Optimization and Deep Reinforcement Learning}

\author{
\IEEEauthorblockN{Yubo Zhang, Hassan ZivariFard, and Xiaodong Wang}\\
\thanks{The authors are with the Department of Electrical Engineering, Columbia University, New York, NY 10027. E-mails: \{yz4891,hz2863, xw2008\}@columbia.edu.}
}
\maketitle
\date{}


\begin{abstract}
\label{sec:Abstract}
We aim to achieve keyless covert communication with a positive-rate in Rayleigh block-fading channels. Specifically, the transmitter and the legitimate receiver are assumed to have either causal or non-causal knowledge of the \ac{CSI} for both the legitimate and the warden channels, while the warden only knows the statistical distribution of the \ac{CSI}. Two problem formulations are considered in this work: (a) Power allocation: maximizing the sum covert rate subject to a maximum power constraint, and (b) Rate allocation: minimizing the power consumption subject to a minimum covert rate constraint. Both problems are formulated based on recent information theoretical results on covert communication over state-dependent channels. When the \ac{CSI} of each fading block is known non-causally, we propose a novel three-step method to solve both the power and rate allocation problems. In the case where the \ac{CSI} is known causally, the power allocation problem can be formulated as \ac{MDP} and be solved using a \ac{DDQN} approach. Although the rate allocation problem under causal \ac{CSI} does not directly conform to an \ac{MDP} structure, it can be approximately solved using the \ac{DDQN} trained for power allocation. Simulation results demonstrate the effectiveness of the proposed power and rate allocation methods and provide comprehensive performance comparisons across different allocation schemes.
\end{abstract}


\section{Introduction} \label{sec:intro}
\IEEEPARstart{W}{ith} the rapid advancement of next-generation wireless systems, particularly the upcoming \ac{6G} networks, security-related challenges have become increasingly critical and are receiving unprecedented attention \cite{BlochBarros}. Traditionally, secure information transmission was guarded by \ac{PLS} techniques \cite{mukherjee2014principles,zou2016survey}, which aim to improve the transmission throughput while ensuring the confidentiality of information. However, in many practical scenarios, it is not only the content of the communication but also its presence that needs to be concealed. To address this concern, covert communication has emerged as a promising paradigm, aiming to hide the act of communication itself from potential adversaries \cite{Bash13,CheISIT13,Bloch16,Wang16}.

\subsection{Information Theoretic Results on Positive-rate Covert Communications} \label{posi_rate_covert_comm}
It is well known that, over a point-to-point channel, only $\mathcal{O}(\sqrt{n})$ bits can be covertly and reliably transmitted in $n$ channel uses. As a result, the covert rate per channel use approaches zero as $n \to \infty$ \cite{Bash13}. To overcome this limitation and achieve a positive covert rate, several enhanced schemes have been proposed. The studies in \cite{Deniable_ITW14,Goeckel15,he2017covert,hayashi2023covert} leveraged noise uncertainty from the warden's perspective, allowing $\mathcal{O}(n)$ bits to be covertly and reliably transmitted over $n$ channel uses. Meanwhile, the works in \cite{bash2016covert,lu2022covert} showed that a positive covert rate can be achieved when the warden lacks precise knowledge of the transmission time. Furthermore, the application of an uninformed jammer \cite{Sobers17,li2020optimal,ISIT22} or a cooperative jammer \cite{ISIT21,ISIT22} can also enable positive-rate covert communications. In addition, it is shown that the knowledge of the channel state information (CSI) and the knowledge of the action-dependent states can help achieve a positive covert rate \cite{LeeWang18,Keyless22,Action_Covert}. In \cite{bendary2021achieving}, a positive covert capacity is achieved over \ac{MIMO} \ac{AWGN} channels. Other approaches to achieving a positive covert rate include employing a full-duplex receiver to generate self-interference that impairs the warden's detection capabilities \cite{shahzad2018achieving}, adopting multi-hop routing to obscure the transmission source \cite{sheikholeslami2018multi}, or randomly activating overt users to confuse the warden \cite{kang2025achieving}. 

\subsection{Optimization of Covert Communications} \label{fad_chan}

Recently, the analysis and optimization of covert communications have attracted significant attention. In \cite{hu2019covert2}, channel inversion power control was adopted under imperfect \ac{CSI}, where the fixed received power was optimized to maximize covert throughput. Covert communication over quasi-static fading channels with unknown channel coefficients and a radiometer-based warden was studied in \cite{shahzad2020covert}, where the transmit power and symbol length were optimized. In \cite{shmuel2021multi}, covert communication over fading channels with a friendly multi-antenna artificial-noise jammer was investigated. Furthermore, a jamming leverage strategy using a neutral node for controlled interference and joint covert beamforming design was proposed in \cite{wang2025use}. Furthermore, \cite{wen2025generative} employed a generative adversarial network to optimize cooperative jamming power in cognitive covert radio networks.

In this paper, unlike \cite{hu2019covert2,shahzad2020covert,shmuel2021multi,lv2021covert}, we study the covert communication over a point-to-point block-fading channel based on the information theoretic results, and we do not assume the existence of a jammer or relay to assist the transmitter. Motivated by the substantial body of work on rate-maximization \cite{hu2019covert2,shahzad2020covert} and power-minimization \cite{lee2008energy,wang2021energy} in block-fading channels, we formulate and tackle both problems in the context of covert communications in this paper.

\subsection{Contributions and Outline} \label{contri_outline}
Of particular relevance to this paper, \cite{Keyless22} characterizes the covert capacity of \acp{DMC} when the \ac{CSI} is available either non-causally or causally at both the transmitter and the receiver. In this paper, we study covert communication with positive rates over block-fading channels, building on the results presented in \cite{Keyless22}. To the best of our knowledge, the present paper is the first to analyze and optimize keyless covert communications over fading channels. The main contributions of this paper are summarized as follows:  
\begin{itemize}
    \item Building on the results of \cite{Keyless22}, we formulate non-convex power allocation optimization problems for achieving maximum positive covert rates with a fixed power budget, as well as non-convex rate allocation optimization problems for achieving a fixed positive covert rate with minimum power consumption, under both non-causal and causal \ac{CSI} at the transmitter.
    \item When the \ac{CSI} is known non-causally at the transmitter, a novel three-step method is proposed to solve the power allocation problem, and another fundamentally different three-step method to solve the rate allocation problem. Extensive simulation results are provided to illustrate the effectiveness of the proposed approaches in solving the non-convex power/rate allocation problems.  
    \item When the \ac{CSI} is known causally at the transmitter, the corresponding power allocation problem is formulated as a \ac{MDP} and solved using a \ac{DRL} approach. On the other hand, the rate allocation problem is not Markovian, thereby is approximately transformed into a power allocation problem and then solved using the aforementioned \ac{DRL} method. 
\end{itemize}


The remainder of this paper is organized as follows. In Section~\ref{sec:sysmodel}, we introduce the system model and problem formulations. In Section~\ref{nc_allo}, we develop algorithms to solve the power and rate allocation problems when the \ac{CSI} is known non-causally at the legitimate terminals. In Section~\ref{c_allo}, we develop \ac{DRL} methods to solve the causal versions of the power and rate allocation problems. Simulation results are presented in Section~\ref{sec:num_result}, and Section~\ref{sec:conclu} concludes the paper. 



\section{System Model and Problem Formulations} \label{sec:sysmodel}
\begin{figure}[t]
\centering
\includegraphics[width=8.3cm]{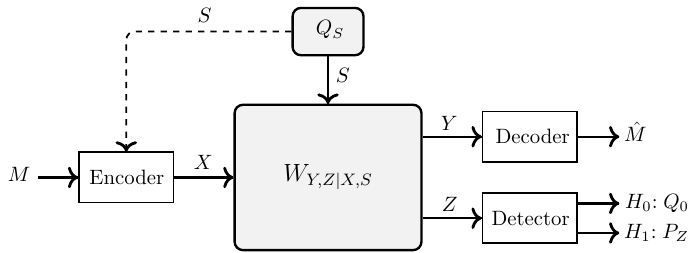}
\caption{Covert communication over \acp{DMC}.}
\label{fig:system_model}
\end{figure}%
\subsection{System Model}
\label{sec:sysmod}
Consider the problem of covert communication over block-fading channels with a channel coherence interval of $T$ symbols. The received signals at the legitimate receiver and the warden, respectively, are
\begin{subequations}\label{block_fad_basis}
\begin{align} 
Y_\ell(t)&=H_\ell X_\ell(t)+N_\ell(t),\label{block_fad_basis_1}\\
Z_\ell(t)&=G_\ell X_\ell(t)+V_\ell(t),\,\,t\in[T]\,\,\text{and}\,\,\ell\in[L],\label{block_fad_basis_2}
\end{align}where we assume the length of the codeword is $LT$, i.e., the codeword spans over $L$ coherence blocks; $X_\ell(t)$ is the $t^{\text{th}}$ symbol during the $\ell^{\text{th}}$ coherence block; $H_\ell \in \mathbb{C}$ and $G_\ell \in \mathbb{C}$ denote the legitimate receiver's and the warden's channel gains, respectively, during the $\ell^{\text{th}}$ coherence block, that are \ac{CSCG} random variables; $N_\ell(t)\sim\calC\calN(0,\sigma_n^2)$ and $V_\ell(t)\sim\calC\calN(0,\sigma_v^2)$ are the additive noise samples at the legitimate receiver and the warden, respectively.
\end{subequations}
Our goal is to design a transmission scheme to maximize the rate at the legitimate receiver while keeping the warden unaware of the transmission.
\subsection{Problem Formulation Based on Information Theoretic Results}
\label{sec:problem_Formu}
Fig.~\ref{fig:system_model} is a schematic diagram of covert communication over \ac{DMC}, when the channel state is available at both the transmitter and the receiver but not at the warden, which is studied in \cite[Sec.~IV]{Keyless22}. 
In Fig.~\ref{fig:system_model}, $M$ is the transmitter's message, $X\in\mathcal{X}$ is the channel input, $Y\in\mathcal{Y}$ and $Z\in\mathcal{Z}$ are the channel outputs of the legitimate receiver and the warden, respectively, $S\in\mathcal{S}$ is the channel state, and $W_{Y,Z|X,S}$ is the channel transition probability. Here, the channel state is \ac{iid} and drawn according to $Q_S$ and we let $x_0\in\mathcal{X}$ be an ``innocent'' symbol corresponding to the absence of communication with the receiver. The distribution induced at the warden in the absence of communication is $Q_0(\cdot) = \sum_{s\in\mathcal{S}} Q_S(s)W_{Z|X,S}(\cdot|x_0,s)$. On the other hand, the distribution induced at the output of the warden by the code design is $P_Z(z)=\sum_{s\in\calS}\sum_{x\in\calX}\sum_{Y\in\calY}Q_S(s)P_{X|S}(x|s)W_{Y,Z|S,X}(y,z|x,s)$. 
The legitimate receiver outputs the decoded message $\hat{M}$ based on the channel output $Y$; whereas the warden attempts to detect the communication by performing a binary hypothesis test between the two distributions $Q_0^{\otimes n}$ and $P_{Z^n}$, corresponding to the absence and presence of communication, respectively. In covert communication, the transmitter aims to communicate with the receiver both reliably and covertly. Reliable means that the probability of error $P_e^{(n)} = \bbP(\hat{M}\ne M)$ vanishes when the codeword length $n\to\infty$. Covert means that the warden cannot determine whether communication is happening (hypothesis $H_1$) or not (hypothesis $H_0$). Specifically, the error probability of the optimal hypothesis test by warden is lower bounded by \cite{Bash13}
\begin{align}
    \alpha_n+\beta_n\ge1-\sqrt{\kld{(P_{Z^n}||Q_0^{\otimes n})}},\label{eq:Covertness_n_letter}
\end{align}where $\alpha_n$ is the probability of false alarm (warden deciding $H_1$ when $H_0$ is true), and $\beta_n$ is the probability of missed detection (warden deciding $H_0$ when $H_1$ is true). Therefore, the code is covert if $\kld{(P_{Z^n}||Q_0^{\otimes n})}$ vanishes when $n\to\infty$.
From \cite[Theorem~1]{Keyless22} the following rate is achievable for \acp{DMC},
\begin{subequations}\label{thm:Capacity_FCSI_NC}
\begin{eqnarray} \max_{P_{X|S}} &&I(X;Y|S),\label{objective_func}\\
\textrm{s.t.}&&P_Z=Q_0,\label{Cost_1}\\
& &I(X;Y|S)\geq I(X;Z|S).\label{Cost_2}
\end{eqnarray}
\end{subequations}
To apply the result in \eqref{thm:Capacity_FCSI_NC} to the block-fading channel defined in \eqref{block_fad_basis}, we let $X\leftarrow\{X_\ell(t)\}_{(\ell,t)\in[L]\times[T]}$,  $Y\leftarrow\{Y_\ell(t)\}_{(\ell,t)\in[L]\times[T]}$, $Z\leftarrow\{Z_\ell(t)\}_{(\ell,t)\in[L]\times[T]}$, and $S\leftarrow\{(H_\ell,G_\ell)\}_{\ell\in[L]}$. Assuming that the total power of a codeword is $P_0$, and the channel instantaneous SNRs are $h_\ell \triangleq \frac{|H_\ell|^2}{\sigma_n^2}$ and $g_\ell \triangleq \frac{|G_\ell|^2}{\sigma_v^2}$. We make the following assumptions:
\begin{enumerate}[(i)]
    \item Gaussian signaling: $X_\ell(t)\sim\calC\calN(0,P_\ell)$, for $\ell\in[L]$, with $\sum_{\ell=1}^LP_\ell\le P_0$.
    \item Quantized channel SNRs: the constraint $I(X;Y|S)\geq I(X;Z|S)$ in \eqref{thm:Capacity_FCSI_NC} implies that we can achieve a positive covert rate only when the legitimate receiver's channel is less noisy than the warden's channel, which is obtained based on the assumption that $S$ is a discrete random variable. Therefore, we assume that the channel SNRs $\{h_\ell,g_\ell, \ell \in [L]\}$ are quantized. Such quantization is also needed when we formulate the problem under causal \ac{CSI} as an \ac{MDP}. Since the number of quantization levels can be arbitrarily large, the quantized system well approximates the original system. 
    \item Here, since transmission occurs over $TL$ time slots and the warden observes the channel outputs $\{Z_\ell(t)\}_{(\ell,t)\in[L]\times[T]}$, the covertness constraint described in \eqref{eq:Covertness_n_letter} reduces to ensuring that
\begin{align}
D\!\left(P_{Z_1^T,\ldots,Z_L^T}\big|\big|Q_0\right)=0, \label{eq:covertness_New}
\end{align}
where $Z_\ell^T=\big(Z_\ell(1),Z_\ell(2),\ldots,Z_\ell(T)\big)$ and $Q_0 = q_0^{\otimes TL}$, where assuming the innocent symbol $x_0=0$, which corresponds to the symbol transmitted by the sender in the no-communication mode, the resulting distribution at the warden’s receiver in this mode is $q_0\sim\mathcal{CN}(0,\sigma_v^2)$.
The covertness constraint in \eqref{eq:covertness_New} therefore restricts the transmitter to communicate only when $G_\ell\approx0$ and $H_\ell>0$.
To derive a more general result and achieve a higher covert rate, inspired by \eqref{eq:Covertness_n_letter}, we relax the strict constraint in \eqref{eq:covertness_New} to the following approximate version:
\begin{align}
D\!\left(P_{Z_1^T,\ldots,Z_L^T}\big|\big|q_0^{\otimes TL}\right)\le\delta, \label{covertness_relaxed}
\end{align}where $\delta\ge0$ quantifies the allowable level of statistical distinguishability between the communication and no-communication modes.
\end{enumerate}
Then the mutual information terms in \eqref{thm:Capacity_FCSI_NC} become $I(X;Y|S)=\sum_{\ell=1}^L \log\left(1+h_\ell P_\ell\right)$ and $I(X;Z|S)=\sum_{\ell=1}^L \log\left(1+g_\ell P_\ell\right)$. Also, the relative entropy in the covertness constraint \eqref{covertness_relaxed} can be written as
\begin{align}
&D\left(P_{Z_1^T...Z_L^T}\big|\big|q_0^{\otimes TL}\right) \nonumber\\
&\qquad\mathop=\limits^{(a)} \sum\limits_{\ell=1}^L D\left(P_{Z_\ell^T}\big|\big|q_0^{\otimes T}\right)\nonumber\\
&\qquad\mathop=\limits^{(b)}\sum\limits_{\ell=1}^LT\cdot D\left(\mathcal{CN}\left(0,\frac{g_\ell P_\ell}{T}+\sigma_v^2\right)\Big|\Big|\,\mathcal{CN}\left(0,\sigma_v^2\right)\right)\nonumber\\
&\qquad\mathop=\limits^{(c)}\sum\limits_{\ell=1}^LT\left(\alpha_\ell-\ln(1+\alpha_\ell)\right)\nonumber\\ 
&\qquad\mathop\le\limits^{(d)}\sum\limits_{\ell=1}^L\frac{T\alpha_\ell^2}{2}\nonumber\\
&\qquad=\frac{1}{2T\sigma_v^4}\sum\limits_{\ell=1}^L(g_\ell P_\ell)^2\nonumber\\
&\qquad\mathop\le\limits^{(e)}\frac{1}{2T\sigma_v^4}\left(\sum\limits_{\ell=1}^Lg_\ell P_\ell\right)^2
,\label{eq:Total_KLD}
\end{align}where
\begin{itemize}
    \item[$(a)$] follows since the channel inputs in \eqref{block_fad_basis} are independent for any time slot $t\in[T]$ and for any block $\ell\in[L]$, and since the additive noise at each time slot is also \ac{iid} and therefore for any block $\ell$, $Z_\ell^T$ is independent of  $\left(Z_{\ell-1}^T,Z_{\ell-2}^T,\cdots,Z_1^T\right)$;
    \item[$(b)$] follows  since, for each $\ell\in[T]$, $X_\ell(1)$, $X_\ell(2)$, $\cdots$, $X_\ell(T)$ are \ac{iid} according to $\mathcal{CN}\left(0,\frac{P_\ell}{T}\right)$, therefore, $Z_\ell(1)$, $Z_\ell(2)$, $\cdots$, $Z_\ell(T)$ are also \ac{iid} according to $\mathcal{CN}\left(0,\frac{g_\ell P_\ell}{T}+\sigma_v^2\right)$;
    \item[$(c)$] follows by defining $\alpha_\ell\triangleq\frac{g_\ell P_\ell}{T\sigma_v^2}$;
    \item[$(d)$] follows since $x-\ln(1+x)\le\frac{x^2}{2}$, for $x\ge0$;
    \item[$(e)$] follows since $g_\ell P_\ell\ge0$, for any $\ell\in[L]$.
\end{itemize}Therefore, to satisfy \eqref{covertness_relaxed}, we can impose that $\sum\limits_{\ell=1}^Lg_\ell P_\ell\le\epsilon_\delta$, where  $\epsilon_\delta \triangleq \sqrt{2T\sigma_v^4\delta}$.
Hence the achievable rate in \eqref{thm:Capacity_FCSI_NC} leads to the following power allocation problem.
\begin{subequations}\label{formu_nc_nonconvex}
\begin{eqnarray} \max_{P_1,\dots,P_L} &&\sum\limits_{\ell=1}^L \log\left(1+h_\ell P_\ell \right),\label{formu_nc_nonconvex_obj}\\
\textrm{s.t.}&&\sum\limits_{\ell=1}^L P_\ell \le P_0,\label{formu_nc_nonconvex_AP}\\
&& \sum\limits_{\ell=1}^L g_\ell P_\ell \leq \epsilon_{\delta}, \label{formu_nc_nonconvex_PBC}\\
&&\sum\limits_{\ell=1}^L \log\left(1+h_\ell P_\ell\right)\ge \sum\limits_{\ell=1}^L \log\left(1+g_\ell P_\ell \right).\label{formu_nc_nonconvex_CCA}
\end{eqnarray}
\end{subequations}
Note that the constraint in \eqref{Cost_2} implies that a positive covert communication rate can be achieved without relying on an external secret key only if the warden’s channel is stochastically degraded with respect to the legitimate receiver’s channel. Motivated by this observation—and since we also do not assume a secret key between the legitimate terminals, we include the constraint \eqref{formu_nc_nonconvex_CCA} in our optimization problem.
The above formulation maximizes the achievable covert rate under a maximum power constraint. Formulation \eqref{formu_nc_nonconvex} is a non-convex optimization problem due to the less noisy constraint \eqref{formu_nc_nonconvex_CCA}. An alternative formulation is to minimize the total transmit power subject to a minimum covert rate constraint. In particular, let $R_\ell = \log(1+ h_\ell P_\ell)$. Then $P_\ell = \frac{e^{R_\ell}-1}{h_\ell}$ and we can convert \eqref{formu_nc_nonconvex} to the following rate allocation problem
\begin{subequations}\label{formu_nonconvex_bit_allo}
\begin{eqnarray} \min_{R_1,\dots,R_L}&&\sum\limits_{\ell=1}^{L} \frac{e^{R_\ell}-1}{h_\ell},\label{formu_bit_object}\\
\textrm{s.t.}&&\sum\limits_{\ell=1}^{L} R_\ell \ge R_0,\label{formu_bit_object_2}\\
&&\sum_{\ell=1}^{L} (e^{R_\ell}-1) \frac{g_\ell}{h_\ell} \leq \epsilon_\delta, \label{formu_bit_object_3}\\
&&\sum_{\ell=1}^{L} R_\ell \ \ge \ \sum_{\ell=1}^{L} \log\left((e^{R_\ell}-1)\frac{g_\ell}{h_\ell}+1\right).\label{less_Noisy_Bit_NC}
\end{eqnarray}
\end{subequations}
Here \eqref{formu_bit_object} and \eqref{formu_bit_object_2} correspond to the minimum total power objective and the minimum covert rate constraint, respectively; \eqref{formu_bit_object_3}  and \eqref{less_Noisy_Bit_NC} are the same covertness constraints as \eqref{formu_nc_nonconvex_PBC} and \eqref{formu_nc_nonconvex_CCA}, respectively.

In the following sections, we will solve the above two formulations under the following two scenarios:
\begin{enumerate}[(i)]
    \item Non-causal \ac{CSI}: $\{(H_\ell,G_\ell)\}_{\ell\in[L]}$ is available at the beginning of the first block. Then by solving the optimization problem in \eqref{formu_nc_nonconvex} or \eqref{formu_nonconvex_bit_allo}, we obtain the power or rate allocation for the entire codeword over $L$ coherence blocks. Note that another scenario that fits such non-causal \ac{CSI} is the parallel channel setup, where $\ell$ denotes the sub-channel (e.g., OFDM subcarrier) of the same coherence block. 
    \item Causal \ac{CSI}: In this case, at the beginning of the $\ell^{\text{th}}$ block, for $\ell\in[L]$, only $\{(H_1,G_1),\dots,(H_\ell,G_\ell)\}$ is available, and we need to determine the power allocation $P_\ell$ and rate allocation $R_\ell$ sequentially.
\end{enumerate}

\section{Power and Rate Allocations with Non-causal CSI}\label{nc_allo}
We assume that the channel gains are known non-causally at both the transmitter and the legitimate receiver and solve the two non-convex optimization problems in \eqref{formu_nc_nonconvex} and \eqref{formu_nonconvex_bit_allo}.

\subsection{Power Allocation Under Non-causal CSI} \label{power_allo_non_causal}
The solution to the power allocation problem in \eqref{formu_nc_nonconvex} consists of three steps: 
\begin{enumerate}[(1)]
    \item First we check the feasibility of achieving positive covert rate.
    \item If the problem is feasible, then we will drop the non-convex less noisy constraint and solve the resulting convex problem. If the solution satisfies the less noisy constraint, then it is the optimal solution to the original problem.
    \item Otherwise we convert  the less noisy constraint to a penalty term and add it to the objective function, and optimize it using the \ac{PGA} method, starting from the convex solution in step (2).
\end{enumerate}

\subsubsection{Feasibility check} \label{feasi_check_power_non}
The non-convex optimization problem in \eqref{formu_nc_nonconvex} has a positive solution if and only if 
\begin{align} \label{power_feasi_condi}
\exists \ell \in [L]: \ h_\ell \ge g_\ell.
\end{align}
For sufficiency, denote $\mathcal{L} = \{\ell \in [L]: h_{\ell} \ge g_\ell \}$, the following solution satisfies all constraints in \eqref{formu_nc_nonconvex_AP}, \eqref{formu_nc_nonconvex_PBC} and \eqref{formu_nc_nonconvex_CCA}: 
\begin{align} \label{suffi_p_allo_def} 
P_\ell = 
\begin{cases} 
\frac{\alpha \epsilon_\delta}{|\mathcal{L}|g_\ell}, & \ \ell \in \mathcal{L}, \\
0, & \ \ell \not \in \mathcal{L},
\end{cases}
 \ \  \text{with} \ \alpha = \min\left\{1, \frac{P_0}{\sum_{\ell \in \mathcal{L}} \frac{\epsilon_\delta}{|\mathcal{L}|g_\ell}} \right\}.
\end{align}
On the other hand, to see the necessity, suppose that $h_\ell < g_\ell$, $\forall \ell \in [L]$. Then to meet the constraint in \eqref{formu_nc_nonconvex_CCA} we have $P_\ell=0$, $\forall \ell \in [L]$, and hence the covert rate is zero.

\subsubsection{Convex Power Allocation} \label{convex_power_solu_non}
If problem is feasible, we first drop the less noisy constraint \eqref{formu_nc_nonconvex_CCA} and solve the resulting convex problem. In particular, we form the Lagrangian
\begin{multline} \label{KKT_formu_B_L_func}
\text{$\mathcal{L}(\bm{P},\lambda,\mu) = \sum_{\ell=1}^L \log\left(1+h_\ell P_\ell\right) 
- \lambda\left(\sum_{\ell=1}^L P_\ell - P_0 \right)$} \\
\text{$\quad - \mu \bigg(\sum_{\ell=1}^L g_\ell P_\ell-\epsilon_{\delta} \bigg),$}
\end{multline}
where $\bm{P}=[P_1,\dots,P_L]$. Since $P_\ell \geq 0$, $\ell \in [L]$, we write down the \ac{KKT} conditions as follows:
\begin{subequations} \label{formu_nc_B}
\begin{align} 
\label{KKT_formu_B_deriv}
&\frac{\partial \mathcal{L}(\bm{P},\lambda,\mu)}{\partial P_\ell} \leq 0 \ \Rightarrow \ P_\ell \geq \frac{1}{\lambda + \mu g_\ell} - \frac{1}{h_\ell}, \ \ell\in[L], \\
\label{KKT_formu_B_deriv_var_product}
&\left(\frac{\partial \mathcal{L}(\bm{P},\lambda,\mu)}{\partial P_\ell}\right) P_\ell = 0 \notag \\
&\Rightarrow \ \left(\frac{1}{P_\ell+\frac{1}{h_\ell}} - \lambda - \mu g_\ell \right) P_\ell = 0, \ \ell\in[L], \\
\label{KKT_formu_B_AP_constraint}
&\sum_{\ell=1}^L P_\ell -P_0 \leq 0, \ \lambda \ge 0, \ \lambda \left(\sum_{\ell=1}^L P_\ell - P_0\right) = 0, \\
\label{KKT_formu_B_slackness}
&\sum_{\ell=1}^L g_\ell P_\ell-\epsilon_{\delta} \leq 0, \ \mu \ge 0, \ \mu\bigg(\sum_{\ell=1}^L g_\ell P_\ell-\epsilon_{\delta}\bigg)=0.
\end{align}
\end{subequations}
If $P_\ell = 0$, from \eqref{KKT_formu_B_deriv}, $\frac{1}{\lambda+\mu g_\ell} \leq \frac{1}{h_\ell}$. On the other hand, if $P_\ell > 0$, from \eqref{KKT_formu_B_deriv_var_product}, $P_\ell = \frac{1}{\lambda+\mu g_\ell} - \frac{1}{h_\ell}$. Therefore
\begin{align} \label{convex_power_solu_1}
P_\ell(\lambda,\mu) = \left(\frac{1}{\lambda+\mu g_\ell} - \frac{1}{h_\ell}\right)^{+}, \ \ell \in [L].
\end{align}

Note that $P_\ell(\lambda,\mu)$ is strictly decreasing in both $\lambda$ and $\mu$. We seek the optimal $(\lambda^*,\mu^*)$ such that the complementary-slackness conditions in \eqref{KKT_formu_B_AP_constraint} and \eqref{KKT_formu_B_slackness} hold. Note that $P_\ell(\lambda, \mu)=0$ holds $\forall \lambda \geq 0$ if $\mu > \mu_\text{max} \triangleq \max_{\ell \in [L]} \frac{h_\ell}{g_\ell}$, hence we set $\mu \in [0, \mu_\text{max}]$. To begin with, we fix $\mu$ and let $\lambda_\text{max}(\mu) = \max_{\ell \in [L]} (h_\ell - \mu g_\ell) \geq 0$, so that $\sum_{\ell=1}^L P_\ell(\lambda_{\text{max}}(\mu),\mu) = 0$ by \eqref{convex_power_solu_1}. If $\sum_{\ell=1}^L P_\ell(0,\mu) < P_0$, then set $\lambda(\mu)=0$ and the total power is not fully used; otherwise there exists a unique $\lambda(\mu) \in [0,\lambda_\text{max}(\mu)]$ satisfying $\sum_{\ell=1}^L P_\ell(\lambda(\mu),\mu) = P_0$, which can be obtained by a bisection search. 


Next we define $f(\lambda(\mu),\mu) \triangleq \sum_{\ell=1}^L g_\ell P_\ell(\lambda(\mu))$, and show that $f(\lambda(\mu),\mu)$ is decreasing in $\mu > 0$. From \eqref{convex_power_solu_1} we have $\sum_{\ell=1}^L P_\ell(0, \mu)=0$ if $\mu > \mu_\text{max}$, and $\sum_{\ell=1}^L P_\ell(0, \mu) \to +\infty$ as $\mu \to 0^{+}$. Therefore for any given $P_0$, there exists $\mu_s \in [0, \mu_\text{max}]$, such that $\sum_{\ell=1}^L P_\ell(0,\mu_s)=P_0$. Then $\forall \mu \in (0,\mu_s)$, since $\sum_{\ell=1}^L P_\ell(0,\mu)>P_0$, from the preceding paragraph, there exists a $\lambda(\mu)$ such that $\sum_{\ell=1}^L P_\ell(\lambda(\mu),\mu) = P_0$. By \eqref{convex_power_solu_1}, we have
\begin{align} \label{f_deriv_mu}
\frac{\partial P_\ell}{\partial \mu} = - \frac{\frac{d \lambda}{d \mu}+g_\ell}{w_\ell} \cdot \mathbbm{1}_{\{P_\ell > 0\}}, \ w_\ell = (\lambda+\mu g_\ell)^2 > 0.
\end{align}
Since $\sum_{\ell=1}^L P_\ell(\lambda(\mu),\mu) = P_0$, we have
\begin{align} \label{sum_power_deriv_mu}
\sum_{\ell=1}^L \frac{\partial P_\ell}{\partial \mu}=0 \Rightarrow \frac{d \lambda}{d \mu} = -\frac{\sum_{\ell=1}^L \frac{g_\ell \mathbbm{1}_{\{P_\ell > 0\}}}{w_\ell}}{\sum_{\ell=1}^L \frac{\mathbbm{1}_{\{P_\ell > 0\}}}{w_\ell}}.
\end{align}
Combining \eqref{f_deriv_mu} - \eqref{sum_power_deriv_mu}, we have
\begin{align} \label{deriv_T_mu}
&\frac{d f(\lambda(\mu),\mu)}{d \mu} = \sum_{\ell=1}^L g_\ell \frac{\partial P_\ell}{\partial \mu} \notag \\
&= \frac{(\sum_{\ell=1}^L \frac{g_\ell \mathbbm{1}_{\{P_\ell > 0\}}}{w_\ell})^2}{\sum_{\ell=1}^L \frac{\mathbbm{1}_{\{P_\ell > 0\}}}{w_\ell}} - \sum_{\ell=1}^L \frac{g_\ell^2 \mathbbm{1}_{\{P_\ell > 0\}}}{w_\ell} \leq 0,
\end{align}
where the last inequality is due to the Cauchy-Schwarz inequality. On the other hand, $\forall \mu \in (\mu_s,\mu_{\text{max}})$, since $\sum_{\ell=1}^L P_\ell(0,\mu)<P_0$, we have $\lambda(\mu)=0$; since each $P_\ell(0,\mu)$ is decreasing in $\mu$ by \eqref{convex_power_solu_1}, so is $f(\lambda(\mu),\mu)$. Therefore $f(\lambda(\mu),\mu)$ is decreasing in $[0,\mu_{\text{max}}]$. If $f(\lambda(0),0) \leq \epsilon_\delta$, $\mu=0$; otherwise we can perform an outer bisection search to find $\mu^* \in [0,\mu_{\text{max}}]$ such that $f(\lambda(\mu^*),\mu^*)=\epsilon_\delta$, and then $\bm{P}^{\text{KKT}} = \{P_\ell(\lambda(\mu^*),\mu^*)\}_{\ell=1}^L$. If $\bm{P}^{\text{KKT}}$ satisfies \eqref{formu_nc_nonconvex_CCA}, then it is the optimal solution to \eqref{formu_nc_nonconvex}. Otherwise, in the last step we use the \ac{PGA} method to solve \eqref{formu_nc_nonconvex}, starting from $\bm{P}^{\text{KKT}}$.

\subsubsection{Projected gradient ascent (PGA)} \label{nonconvex_power_solu_non}
Define 
\begin{align} \label{delta_def}
\Delta(\bm{P}) = \sum_{\ell=1}^L \log\left(1+P_\ell h_\ell\right)-\sum_{\ell=1}^L \log\left(1+P_\ell g_\ell\right).
\end{align}
Then the less noisy constraint in \eqref{formu_nc_nonconvex_CCA} becomes $\Delta(\bm{P}) \geq 0$. 

Using the penalty method, we convert the convex constraint in \eqref{formu_nc_nonconvex_CCA} to a penalty term to form the following penalized objective function to be maximized
\begin{align} \label{obj_func_penalty_term}
\mathcal{F}(\bm{P},\eta,b) = \sum_{\ell=1}^L \log\left(1+P_\ell h_\ell \right) - \eta \bigg[\Delta(\bm{P}) - b \bigg]^2,
\end{align}
where $\eta > 0$ is the penalty factor and $b \ge 0$ is an auxiliary variable. 

The constraints in \eqref{formu_nc_nonconvex_AP} and \eqref{formu_nc_nonconvex_PBC} form a convex feasible set
\begin{align} \label{power_fea_set}
\mathcal{P} = \left\{\bm{P} \ \Bigg| \ \sum_{\ell=1}^L P_{\ell} \leq P_0, \ \sum_{\ell=1}^L g_\ell P_{\ell} \leq \epsilon_{\delta} \right\}.
\end{align}

In general, to ensure that the original objective function is sufficiently maximized, the penalty factor $\eta$ should be initialized with a large value. Then it will be gradually increased to guarantee that the penalty term approaches zero. In our case, however, if we initialize as $\bm{P}^{(0)} = \bm{P}^{\text{KKT}}$, $\ell \in [L]$, and $b^{(0)}=0$, then effectively we start from the maximum of the original objective, which is infeasible, i.e., $\Delta(\bm{P}^{\text{KKT}}) < 0$. Hence we can initialize $\eta$ as 
\begin{align} \label{lam_origin_set}
\eta_0 = \frac{C}{\Delta^2(\bm{P}^{(0)})} \ \sum_{\ell=1}^L \log\left(1+P_\ell^{(0)} h_\ell\right),
\end{align}
where $C \gg 1$ is a constant, so that the initial penalized objective is $(1-C) \sum_{\ell=1}^L \log\left(1+P_\ell^{(0)} h_\ell\right) \ll 0$.
Since the original problem in \eqref{formu_nc_nonconvex} is feasible, there is a solution in $\mathcal{P}$ with positive penalized objective. Then we apply the following \ac{PGA} procedure: for $n=0,1,\dots,$
\begin{align} \label{proj_gd_define}
\bm{P}^{(n+1)} &= \Pi_{\mathcal{P}} \bigg[ \bm{P}^{(n)} + \alpha_1 \nabla_{\bm{P}} \mathcal{F} \bigg] \bigg|_{\bm{P}=\bm{P}^{(n)}}, \notag \\
b^{(n+1)} &= \Pi_{\mathbb{R}^{+}} \bigg[ b^{(n)} + \alpha_2 \frac{\partial \mathcal{F}}{\partial b} \bigg] \bigg|_{b=b^{(n)}},
\end{align}
where $\alpha_1$ and $\alpha_2$ are step sizes, $\Pi_{\mathbb{R}^{+}}(b) = \max\{b,0\}$, and $\Pi_{\mathcal{P}}$ is a projector onto the feasible set $\mathcal{P}$ in \eqref{power_fea_set}, given by
\begin{align} \label{pi_p_projector}
\Pi_{\mathcal{P}}(\bm{P}) = \frac{\bm{P}}{\max\left\{1, \frac{\sum_{\ell=1}^L P_\ell}{P_0}, \frac{\sum_{\ell=1}^L g_\ell P_\ell}{\epsilon_{\delta}} \right\}}.
\end{align}
The penalty factor $\eta$ in \eqref{obj_func_penalty_term} follows a linear growth and is updated every $N$ iterations of \ac{PGA}. Specifically, at the $n^{\text{th}}$ update step in \eqref{proj_gd_define},  $\mathcal{F}$ is given by $\mathcal{F}(\bm{P}^{(n)},\eta_n,b^{(n)})$, where $\eta_n$ is set as
\begin{align} \label{lam_growth}
\eta_n = \eta_0 \left(1+\gamma \left\lfloor \frac{n}{N} \right\rfloor \right), 
\end{align}
where $\gamma > 0$ is a constant that controls the growth rate. The \ac{PGA} iteration in \eqref{proj_gd_define} stops when it reaches convergence, i.e.,
\begin{align} \label{power_pga_stop_condi}
\max\bigg\{ \| \bm{P}^{(n)} - \bm{P}^{(n-1)} \|, \ \lvert b^{(n)}-b^{(n-1)} \rvert \bigg\} < \omega.
\end{align}
Upon convergence, if $\Delta(\bm{P}) \geq 0$, then $\bm{P}^{(n)}$ is feasible and output as the solution to \eqref{formu_nc_nonconvex}. Otherwise \ac{PGA} does not find a feasible solution, and \eqref{suffi_p_allo_def} is output as the solution to \eqref{formu_nc_nonconvex}. 

\textbf{Complexity Analysis:} Let the stopping rules for the bisection search in \eqref{convex_power_solu_1} be $\frac{|f(\lambda(\mu),\mu)-\epsilon_\delta|}{\epsilon_\delta} \leq \tau$. Denote by $T_{\omega}$ the number of \ac{PGA} steps to meet the stopping rule in \eqref{power_pga_stop_condi}. The complexity of the feasibility check in \eqref{power_feasi_condi} together with the trivial solution calculation in \eqref{suffi_p_allo_def} is $\mathcal{O}(L)$. Next, for the convex power allocation, the two-level bisection search takes $\mathcal{O}((\log(1 / \tau))^2)$ iterations, and the computational complexity in each iteration is $\mathcal{O}(L)$, hence the overall complexity is $\mathcal{O}(L (\log(1 / \tau))^2)$. Finally, for the \ac{PGA}, since the computations in \eqref{delta_def}, \eqref{lam_origin_set}, and \eqref{proj_gd_define} each has a complexity of $\mathcal{O}(L)$, the overall complexity is $\mathcal{O}(L T_{\omega})$. Hence the overall complexity is $\mathcal{O}(L ((\log(1 / \tau))^2+T_{\omega}))$.

\subsection{Rate Allocation Under Non-causal CSI}\label{bit_allo_non_causal}
Now we consider the rate allocation problem in \eqref{formu_nonconvex_bit_allo}. Similar to Sec.~\ref{power_allo_non_causal}, the solution consists of three steps.

\subsubsection{Infeasibility check} \label{feasi_check_bit}
Unlike \eqref{formu_nc_nonconvex}, for the problem in \eqref{formu_nonconvex_bit_allo}, we only have the following two sufficient conditions for \textit{infeasibility}. First, if condition \eqref{power_feasi_condi} is not satisfied, i.e., 
\begin{align} \label{simple_infeasi_condi}
\forall \ell \in [L]: \ h_\ell < g_\ell,
\end{align}
the covert rate is zero and hence the problem \eqref{formu_nonconvex_bit_allo} is infeasible. 
Second, based on \eqref{formu_bit_object_2} and \eqref{formu_bit_object_3}, we consider the following convex problem
\begin{align} \label{check_min_covert}
\min_{R_1, \dots, R_L} \ \sum_{\ell=1}^{L}(e^{R_\ell}-1) \frac{g_\ell}{h_\ell}, \ \ \text{s.t.} \ \ \sum_{\ell=1}^L R_\ell \geq R_0,
\end{align}
whose solution is given by
\begin{align} \label{wf_solu_check}
R_\ell^{\text{opt}} = \left[\log\left(\frac{\eta^* h_\ell}{g_\ell}\right)\right]^{+}, 
\end{align}
Then if 
\begin{align} \label{convex_rate_infea_condi}
\sum_{\ell=1}^{L}(e^{R_\ell^{\text{opt}}}-1) \frac{g_\ell}{h_\ell} \geq \epsilon_\delta,
\end{align}
we declare that the problem \eqref{formu_nonconvex_bit_allo} is infeasible. If both \eqref{simple_infeasi_condi} and \eqref{convex_rate_infea_condi} do not hold, we drop the less noisy constraint in \eqref{less_Noisy_Bit_NC} and solve the resulting convex rate allocation problem, as follows.

\subsubsection{Convex Rate Allocation} \label{convex_nc_bit_allo}
We form the Lagrangian
\begin{multline} \label{KKT_formu_L_func_bit}
\text{$\mathcal{L}(\bm{R},\lambda,\mu) = \sum_{\ell=1}^L \frac{e^{R_{\ell}}-1}{h_{\ell}} 
- \lambda\left(\sum_{\ell=1}^L R_\ell - R_0 \right)$} \\
\text{$\quad + \mu \left[ \sum_{\ell=1}^L \left(e^{R_{\ell}}-1\right)\frac{g_{\ell}}{h_{\ell}} - \epsilon_\delta \right],$}
\end{multline}
where $\bm{R}=[R_1,\dots,R_L]$. Since $R_\ell \geq 0$, $\ell \in [L]$, we write down the \ac{KKT} conditions as follows:
\begin{subequations} \label{KKT_convex_bit_allo}
\begin{align} 
\label{KKT_formu_deriv_bit}
&\frac{\partial \mathcal{L}(\bm{R},\lambda,\mu)}{\partial R_\ell} \geq 0 \ \Rightarrow \ R_\ell \geq \log\left(\frac{\lambda h_\ell}{1+\mu g_\ell}\right), \ \ell\in[L], \\
\label{KKT_formu_deriv_rate_product}
&\left(\frac{\partial \mathcal{L}(\bm{R},\lambda,\mu)}{\partial R_\ell}\right) R_\ell = 0 \notag  \\
&\Rightarrow \ \left((1+\mu g_\ell) \frac{e^{R_\ell}}{h_\ell} - \lambda \right) R_\ell = 0, \ \ell \in [L], \\
\label{KKT_formu_bit_schedule}
&\sum_{\ell=1}^L R_\ell -R_0 \geq 0, \ \lambda \ge 0, \ \lambda\left( \sum_{\ell=1}^{L} R_\ell - R_0 \right) = 0, \\
\label{KKT_formu_slackness_bit}
&\sum_{\ell=1}^L \left(e^{R_{\ell}}-1\right)\frac{g_{\ell}}{h_{\ell}} -\epsilon_\delta \leq 0, \ \mu \ge 0, \notag \\ 
&\mu \left[ \sum_{\ell=1}^L \left(e^{R_{\ell}}-1\right)\frac{g_{\ell}}{h_{\ell}} -\epsilon_\delta \right]=0. 
\end{align}
\end{subequations}
If $R_\ell = 0$, from \eqref{KKT_formu_deriv_bit} $\lambda h_\ell \leq 1+\mu g_\ell$. On the other hand, if $R_\ell > 0$, from \eqref{KKT_formu_deriv_rate_product} $R_\ell = \log\left(\frac{\lambda h_\ell}{1+\mu g_\ell}\right)$. Therefore, 
\begin{align} \label{convex_rate_solu_1}
R_\ell(\lambda,\mu) = \left[ \log\left(\frac{\lambda h_\ell}{1+\mu g_\ell}\right) \right]^{+}, \ \ell \in [L].
\end{align}
We first fix $\mu$ and let $\lambda_{\text{max}}(\mu) \triangleq \max_{\ell\in[L]} \frac{(1+\mu g_\ell)e^{R_0 / L}}{h_\ell}$, so that $\sum_{\ell=1}^L R_\ell(\lambda_{\text{max}}(\mu),\mu) > R_0$ by \eqref{convex_rate_solu_1}. On the other hand, by \eqref{convex_rate_solu_1} $R_\ell(\lambda,\mu)<R_\ell^{\text{opt}}$ when $\lambda = \eta^* \mu$, hence $\sum_{\ell=1}^L R_\ell(\eta^* \mu,\mu) < \sum_{\ell=1}^L R_\ell^{\text{opt}}=R_0$. Since $R_\ell(\lambda,\mu)$ increases with $\lambda$, there exists a unique $\lambda(\mu) \in [\eta^* \mu, \lambda_{\text{max}}(\mu)]$ satisfying $\sum_{\ell=1}^L R_\ell(\lambda(\mu),\mu) = R_0$, which can be obtained by a bisection search.


Define $g(\lambda(\mu),\mu) \triangleq \sum_{\ell=1}^L \left(e^{R_{\ell}(\lambda(\mu),\mu)}-1\right) \frac{g_{\ell}}{h_{\ell}}$. Similar to \eqref{f_deriv_mu}-\eqref{deriv_T_mu}, one can show that $g(\lambda(\mu),\mu)$ is decreasing in $\mu \geq 0$. Based on \eqref{convex_rate_solu_1}, we have $\lim_{\mu \rightarrow +\infty} \sum_{\ell=1}^L R_\ell = \lim_{\mu \rightarrow +\infty} \sum_{\ell=1}^L \left[ \log\left(\frac{\lambda(\mu)}{\mu} \cdot \frac{h_\ell}{g_\ell}\right) \right]^{+} = R_0$. Comparing this with \eqref{wf_solu_check}, we conclude that $\lim_{\mu \rightarrow +\infty} \ R_\ell(\lambda(\mu),\mu) = R_\ell^{\text{opt}}$, $\ell \in [L]$. 

Since the infeasibility condition \eqref{convex_rate_infea_condi} does not hold, we write 
\begin{align} \label{g_mu_lim}
g^*(\mu) \triangleq \text{LHS of (28)} = \sum_{\ell=1}^{L} \left( \eta^* - \frac{g_\ell}{h_\ell} \right)^{+} = \epsilon_\delta - \kappa,
\end{align}
for some $\kappa > 0$. If $g(\lambda(0),0) \leq \epsilon_\delta$, we set $\mu=0$. Otherwise, since $\lim_{\mu \to +\infty} g(\lambda(\mu),\mu) = g^*(\mu) = \epsilon_\delta-\kappa$, there exists $\mu \in (0,+\infty)$ such that $g(\lambda(\mu),\mu)=\epsilon_\delta$. To perform bisection search, we derive an upper bound $\mu_\text{max}$, such that $g(\lambda(\mu_\text{max}),\mu_\text{max}) < \epsilon_\delta$. Based on \eqref{wf_solu_check} and \eqref{convex_rate_solu_1}, we have
\begin{align} \label{eta_prop_bound}
&\sum_{\ell=1}^{L} \left[ \log \left( \frac{\eta^* h_\ell}{g_\ell} \right) \right]^{+} = R_0 = \sum_{\ell=1}^L R_\ell(\lambda(\mu),\mu) \notag \\
&\geq \sum_{\ell=1}^L \left[\log\left(\frac{\lambda(\mu) h_\ell}{\mu g_\ell \left(1+\frac{1}{\mu g_\text{min}}\right)}\right)\right]^{+},
\end{align}
where $g_\text{min} = \min_{\ell \in [L]} g_\ell$. Then we have $\frac{\lambda(\mu)}{\mu} \leq \eta^* \left(1+\frac{1}{\mu g_\text{min}}\right)$. On the other hand, note that
\begin{align} \label{g_func_bound}
&g(\lambda(\mu),\mu) = \sum_{\ell=1}^L \left(e^{R_{\ell}(\lambda(\mu),\mu)}-1\right) \frac{g_{\ell}}{h_{\ell}} \notag \\
& < \sum_{\ell=1}^L \left(e^{\left[\log\left(\frac{\lambda(\mu) h_\ell}{\mu g_\ell}\right)\right]^{+}}-1\right) \frac{g_{\ell}}{h_{\ell}} = \sum_{\ell=1}^L \left(\frac{\lambda(\mu)}{\mu} - \frac{g_\ell}{h_\ell}\right)^{+} \notag \\
& \leq \sum_{\ell=1}^L \left(\eta^* \left(1+\frac{1}{\mu g_\text{min}}\right) - \frac{g_\ell}{h_\ell}\right)^{+} \notag \\
& \leq \sum_{\ell=1}^L \left(\eta^* - \frac{g_\ell}{h_\ell}\right)^{+} + \frac{\eta^* L}{\mu g_\text{min}} = g^*(\mu) + \frac{\eta^* L}{\mu g_\text{min}}.
\end{align}
Since $g^*(\mu) = \epsilon_\delta - \kappa$, we set $\mu_\text{max} \triangleq \frac{2 \eta^* L}{g_\text{min} \kappa}$, so that
\begin{align} \label{mu_max_feasi}
g(\lambda(\mu_\text{max}),\mu_\text{max}) < \epsilon_\delta - \frac{\kappa}{2}.
\end{align}
Hence by a bisection search over $[0,\mu_\text{max}]$, we obtain $\mu^*$ and $\bm{R}^{\text{KKT}}=\{R_\ell(\lambda(\mu^*),\mu^*)\}_{\ell=1}^L$. If $\bm{R}^{\text{KKT}}$ satisfies the less noisy constraint \eqref{less_Noisy_Bit_NC}, then it is the optimal solution to the original problem in \eqref{formu_nonconvex_bit_allo}. Otherwise, in the last step we apply the \ac{PGD} method to the non-convex optimization problem in \eqref{formu_nonconvex_bit_allo}, starting from $\bm{R}^{\text{KKT}}$.

\subsubsection{Projected gradient descent (PGD)} \label{nonconvex_nc_bit_allo}
Define  
\begin{align} \label{delta_def_rate}
\Delta(\bm{R}) = \sum_{\ell=1}^{L} R_\ell -\sum_{\ell=1}^{L} \log\left( 1+\left(e^{R_\ell}-1\right)\frac{g_\ell}{h_\ell} \right).
\end{align}
Then the less noisy constraint in \eqref{less_Noisy_Bit_NC} becomes $\Delta(\bm{R}) \geq 0$. 

Using the penalty method, we convert the convex constraint in \eqref{less_Noisy_Bit_NC} to a penalty term to form the following penalized objective function to be minimized
\begin{align} \label{rate_func_penalty_term}
\mathcal{G}(\bm{R},\eta,b) = \sum_{\ell=1}^{L} \frac{e^{R_{\ell}}-1}{h_\ell} + \eta \bigg[\Delta(\bm{R}) - b \bigg]^2,
\end{align}
where $\eta > 0$ is the penalty factor and $b \ge 0$ is an auxiliary variable.

The constraints in \eqref{formu_bit_object_2} and \eqref{formu_bit_object_3} form a feasible set: 
\begin{align} \label{rate_fea_set}
\mathcal{B} = \left\{ \bm{R} \ \Bigg| \ \sum_{\ell=1}^L (e^{R_\ell}-1) \frac{g_\ell}{h_\ell} \leq \epsilon_\delta, \ \sum_{\ell=1}^L R_{\ell} \geq R_0 \right\}.
\end{align}

We initialize as $\bm{R}^{(0)} = \bm{R}^{\text{KKT}}$, $\ell \in [L]$, $b^{(0)}=0$, and
\begin{align} \label{rate_lam_origin_set}
\eta_0 = \frac{C}{\Delta^2(\bm{R}^{(0)})} \ \sum_{\ell=1}^L \frac{e^{R_{\ell}}-1}{h_\ell},
\end{align}
where $C \gg 1$ is a constant, so that the initial penalized objective is $(1+C) \sum_{\ell=1}^L \frac{e^{R_{\ell}}-1}{h_\ell} \gg 0$.
To minimize \eqref{rate_func_penalty_term}, we run the following \ac{PGD} procedure: for $n=0,1,\dots,$
\begin{align} \label{rate_proj_gd_define}
\bm{R}^{(n+1)} &= \Pi_{\mathcal{B}} \bigg[ \bm{R}^{(n)} - \alpha_1 \nabla_{\bm{R}} \mathcal{G} \bigg] \bigg|_{\bm{R}=\bm{R}^{(n)}}, \notag \\
b^{(n+1)} &= \Pi_{\mathbb{R}^{+}} \bigg[ b^{(n)} - \alpha_2 \frac{\partial \mathcal{F}}{\partial b} \bigg] \bigg|_{b=b^{(n)}}.
\end{align}
For the projector $\Pi_{\mathcal{B}}$, note that $\mathcal{B}$ is an intersection of two convex sets, i.e., $\mathcal{B} = \mathcal{B}_1 \cap \mathcal{B}_2$, where $\mathcal{B}_1 = \left\{\bm{R} \ \big| \ \sum_{\ell=1}^L (e^{R_\ell} - 1) \frac{g_\ell}{h_\ell} \leq \epsilon_\delta \right\}$ and $\mathcal{B}_2 = \left\{ \bm{R} \ | \ \sum_{\ell=1}^{L} R_\ell \geq R_0 \right\}$.
$\Pi_{\mathcal{B}}(\bm{R})$ can be computed by using the \ac{POCS} algorithm. Specifically, for $i=0,2,4,\dots$, $\bm{R}^{i+1} = \Pi_{\mathcal{B}_1}(\bm{R}^{i})$, where 
\begin{align} \label{pi_b1_proj}
\Pi_{\mathcal{B}_1}(\bm{R}) &= \left[R_\ell + \min\left\{ 0, \log\left(\frac{\epsilon_\delta+B}{A}\right)\right\}, \ell\in[L]\right], \notag \\
A &= \sum_{\ell=1}^L \frac{g_\ell}{h_\ell} e^{R_\ell}, \ B = \sum_{\ell=1}^L \frac{g_\ell}{h_\ell}.
\end{align}
And $\bm{R}^{i+2} = \Pi_{\mathcal{B}_2}(\bm{R}^{i+1})$ with $\Pi_{\mathcal{B}_2}(\bm{R}) = \bm{R} \cdot 1_{\{\sum_{\ell=1}^{L} R_\ell \geq R_0\}} + \bm{R} \frac{R_0}{\sum_{\ell=1}^{L} R_\ell} \cdot 1_{\{\sum_{\ell=1}^{L} R_\ell < R_0\}}$.
The penalty factor $\eta$ is updated every $N$ \ac{PGD} steps according to \eqref{lam_growth}. The iteration in \eqref{rate_proj_gd_define} stops when it reaches convergence, i.e.,
\begin{align} \label{rate_pgd_stop_condi}
\max\bigg\{ \| \bm{R}^{(n)} - \bm{R}^{(n-1)} \|, \ \lvert b^{(n)}-b^{(n-1)} \rvert \bigg\} < \omega.
\end{align}
Upon convergence, if $\Delta(\bm{R}^{(n)}) \geq 0$, then $\bm{R}^{(n)}$ is feasible and output as the solution to \eqref{formu_nonconvex_bit_allo}. Otherwise we consider the power allocation in \eqref{suffi_p_allo_def} and denote it as $\bm{P}(\alpha)$. Denote $R_\ell(\alpha)=\log\left( 1+P_\ell(\alpha) h_\ell \right)$ and $R_{\text{sum}}(\alpha) = \sum_{\ell=1}^L R_\ell(\alpha)$. If $R_{\text{sum}}(1) \geq R_0$, 
then $\bm{R}(1) = \left[R_1(1),\dots,R_L(1)\right]$ is a feasible solution to the original problem in \eqref{formu_nonconvex_bit_allo}. Since $R_{\text{sum}}(\alpha)$ monotonically increases with $\alpha$, and $R_{\text{sum}}(0)=0$, using bisection search we can find a unique $\alpha_0 \in [0,1]$, such that,
\begin{align} \label{R_trivial_final_solu}
\bm{R}(\alpha_0) = \left[R_1(\alpha_0),\dots,R_L(\alpha_0)\right], \ \ \text{s.t.} \ \ R_{\text{sum}}(\alpha_0)=R_0.
\end{align}
Otherwise, if $R_{\text{sum}}(1) < R_0$, we declare that \eqref{formu_nonconvex_bit_allo} is infeasible. Note that for the last case of infeasibility declaration, the original non-convex problem in \eqref{formu_nonconvex_bit_allo} may still be feasible, but the suboptimal \ac{PGD} method cannot find a feasible solution. 

\textbf{Complexity Analysis:} The complexity of the non-causal rate allocation scheme is the same as that of the non-causal power allocation scheme, except for the worst case in \eqref{R_trivial_final_solu}, where the \ac{PGD} does not converge to a feasible solution and a bisection search with tolerance $\tau_{\alpha}$ is needed to find an alternative solution. Under this condition, the complexity is $\mathcal{O}(L ((\log(1 / \tau))^2+\log(1 / \tau_{\alpha})+T_{\omega}))$.

\section{Power and Rate Allocations with causal CSI}\label{c_allo}

In this section, we treat the causal versions of the problems in \eqref{formu_nc_nonconvex} and \eqref{formu_nonconvex_bit_allo}, where at each block $\ell \in [L]$, the power $P_\ell$ and rate $R_\ell$, respectively, are computed sequentially based on the causal channel state information $\{(h_k,g_k), k\in[\ell]\}$. We formulate the corresponding power allocation problem as a \ac{MDP} and use a \ac{DRL} method, in particular, the \ac{DDQN} method, to solve it. The sequential rate allocation is not an \ac{MDP}, but we show that it can be solved approximately using the trained \ac{DDQN} for power allocation.

\subsection{Power Allocation Under Causal CSI}\label{power_c_allo}
\subsubsection{MDP formulation and Q-learning} \label{mdp_formu}
Since the power allocation $P_\ell$ is sequentially computed for each $\ell \in [L]$, the less noisy constraint in \eqref{formu_nc_nonconvex_CCA} is enforced for each $\ell$:
\begin{align} \label{formu_c_CCA}
\sum_{k=1}^{\ell} \log\left(1+P_k h_k\right) \ge \sum_{k=1}^{\ell} \log\left(1+P_k g_k\right), \ \ell \in [L].
\end{align}
We next formulate the sequential power allocation to maximize the sum covert rate in \eqref{formu_nc_nonconvex_obj}, subject to the constraints in \eqref{formu_nc_nonconvex_AP}, \eqref{formu_nc_nonconvex_PBC} and \eqref{formu_c_CCA} as an \ac{MDP}. Recall \eqref{formu_nc_nonconvex_AP}, denote $P^{(\ell)}$ as the remaining power before allocating power to the $\ell^{\text{th}}$ block, and $P_\ell$ as the power allocated to the $\ell^{\text{th}}$ block. Then we have
\begin{align} \label{power_budget_property}
P^{(1)}=P_0, \  P^{(\ell+1)}=P^{(\ell)}-P_\ell, \ \ell \in [L-1].
\end{align}
Similarly, according to the constraint in \eqref{formu_nc_nonconvex_PBC}, denote $\epsilon^{(\ell)}$ as the remaining covertness margin before allocating power to the $\ell^{\text{th}}$ block, we have that
\begin{align} \label{covert_margin_property}
\epsilon^{(1)}=\epsilon_\delta, \  \epsilon^{(\ell+1)}=\epsilon^{(\ell)}-P_\ell g_\ell, \ \ell \in [L-1].
\end{align}
Moreover, define 
\begin{align} \label{cca_budget_property}
\Gamma^{(1)}\triangleq0, \ \Gamma^{(\ell+1)} \triangleq \Gamma^{(\ell)}+\log\left(\frac{1+P_{\ell}h_\ell}{1+P_{\ell}g_\ell}\right), \ \ell \in [L-1]. 
\end{align}
Then the constraints in \eqref{formu_c_CCA} become
\begin{align} \label{cca_constraint_simpler}
\Gamma^{(\ell)} \ge 0, \ \ell \in [L].
\end{align}
In our \ac{MDP} formulation, for each $\ell \in [L]$, the state is defined as $\bm{s}_\ell = (P^{(\ell)},\epsilon^{(\ell)},\Gamma^{(\ell)},h_\ell,g_\ell)$, and the action is $P_\ell$. Recall that the channel SNRs $h_\ell$ and $g_\ell$ are quantized, and similarly we assume that the power allocation $P_\ell \in \mathcal{P} = \{n \Delta_P, n=0,1,2,\dots\}$, where $\Delta_P$ is the quantization step size. Based on \eqref{formu_nc_nonconvex_AP}, \eqref{power_budget_property} and \eqref{formu_nc_nonconvex_PBC}, we have $0 \leq P_\ell \leq P^{(\ell)}$ and $0 \leq P_\ell \leq \frac{\epsilon^{(\ell)}}{g_\ell}$. Furthermore, to meet the constraints in \eqref{cca_constraint_simpler}, we have
\begin{align} \label{gamma_p_constraint}
\Gamma^{(\ell+1)}=\Gamma^{(\ell)}+\log\left(\frac{1+h_\ell P_\ell}{1+g_\ell P_\ell}\right) &\ge 0  \notag \\ 
\Leftrightarrow \ P_\ell\left(e^{-\Gamma^{(\ell)}} g_\ell - h_\ell\right) &\leq 1-e^{-\Gamma^{(\ell)}}. 
\end{align}
Hence the feasible set of the action $P_\ell$ is given by \eqref{feasi_set_power_causal}.
\begin{figure*}[t]
\begin{equation} \label{feasi_set_power_causal}
\mathcal{A}_\ell = \left\{ P \in \mathcal{P} \ \bigg| \ P\leq \min \left\{P^{(\ell)}, \frac{\epsilon^{(\ell)}}{g_\ell}\right\}, \ \text{if} \ e^{-\Gamma^{(\ell)}}g_\ell \leq h_\ell \ \ \& \ \ P\leq \min \left\{P^{(\ell)}, \frac{\epsilon^{(\ell)}}{g_\ell},\frac{1-e^{-\Gamma^{(\ell)}}}{e^{-\Gamma^{(\ell)}}g_\ell - h_\ell}\right\}, \ \text{otherwise} \right\}
\end{equation}
\end{figure*}
At each block $\ell$, an action $P_\ell \in \mathcal{A}_\ell$ is selected based on the current state $\bm{s}_\ell$, according to a policy $\pi(P|\bm{s}) = \mathbb{P}(P_\ell=P | \bm{s}_\ell=\bm{s})$, which defines the probability distribution over actions given the state. The state then transitions to a new one $\bm{s}_{\ell+1}$, yielding a reward $r_{\ell+1}$, which is the covert rate, given by
\begin{align} \label{reward_func_define}
r_{\ell+1} = r(\bm{s}_\ell,P_\ell) \triangleq \log\left(1+h_\ell P_\ell \right). 
\end{align}
These variables together form an \ac{MDP} experience, denoted as $\bm{e} = (\ell,\bm{s}_\ell,P_\ell,\bm{s}_{\ell+1},r_{\ell+1})$. Assuming that $L$ is large, for a given policy $\pi$, the expected cumulative reward function at each block, commonly known as the Q-function, is defined as
\begin{align} \label{q_s_a_def}
Q_{\pi}(&\bm{s},P) = r(\bm{s},P) \notag \\
&+ \mathop{\mathbb{E}}_{P_k \sim \pi(\cdot | \bm{s}_k)} \left[\sum_{k=\ell+1}^{L} r(\bm{s}_{k},P_k) \ \bigg| \ \bm{s}_\ell=\bm{s}, P_\ell=P \right].
\end{align}
The optimal policy $\pi^{*}$ satisfies $Q_{\pi^{*}}(\bm{s},P) \geq Q_{\pi}(\bm{s},P)$ for all state-action pairs $(\bm{s},P)$ and all other policies $\pi$. Note that any finite \ac{MDP} admits an optimal deterministic policy \cite[Theorem 17.8]{mohri_Book}. 

A basic method to find the optimal policy $\pi^{*}$ is Q-learning, an iterative algorithm that learns from episodes of experiences. Suppose that the $j^{\text{th}}$ experience corresponds to block $\ell$. Then given the state $\bm{s}_\ell$, the $\xi$-greedy policy is applied using the current Q-function to select the action $P_\ell$: it takes the greedy action $P_\ell = \argmax_{P \in \mathcal{A}_\ell} Q(\bm{s}_\ell,P)$ with probability $(1-\xi)$ or randomly chooses an action $P_\ell \in \mathcal{A}_\ell$ with probability $\xi$. Then the reward $r_{\ell+1}$ in \eqref{reward_func_define} is obtained and the state transitions to $\bm{s}_{\ell+1}$. The $j^{\text{th}}$ experience $\bm{e}_j = (\ell,\bm{s}_\ell,P_\ell,\bm{s}_{\ell+1},r_{\ell+1})$ is used to update the entry $(\bm{s}_\ell,P_\ell)$ of the Q-function, as follows,
\begin{align} \label{basic_q_learn}
Q(\bm{s}_\ell,P_\ell) \ &\leftarrow \ (1-\beta) Q(\bm{s}_\ell,P_\ell) \notag \\
&+ \beta \left( r_{\ell+1}+\max_{P_{\ell+1} \in \mathcal{A}_{\ell+1}} Q(\bm{s}_{\ell+1},P_{\ell+1}) \right), 
\end{align}
where $\beta \in \left(0,1\right]$ is a learning rate parameter. However, in this case, the cardinality of the state space, i.e., the total number of possible values that the state $\bm{s}_\ell$ can take, is very large. Therefore it is impractical to obtain the Q-function values for all state-action pairs, and we resort to the neural network representation of the Q-function, i.e., the \ac{DQN}.

\subsubsection{DDQN solution} \label{ddqn_solu}
The double deep Q-learning network (DDQN) consists of a primary Q-network $Q(\bm{s}, P; \Theta)$ and a target Q-network $\hat{Q}(\bm{s},P;\hat{\Theta})$, with the input to each Q-network as the state-action pair $(\bm{s}, P)$, and the output as the corresponding Q-function value. The primary Q-network is periodically copied into the target Q-network every $N_{\text{ep}}$ training episodes to ensure more stable updates (line 23 of Alg.~\ref{alg_3}). At each training episode, $N_{\text{tr}}$ $L$-state transitions are carried out based on the current primary network $Q(\bm{s},P;\Theta)$, and the corresponding experiences are stored in the replay buffer $\mathcal{D}$ of size $N_{\text{buff}}$ (lines 6-18 of Alg.~\ref{alg_3}). Specifically, similar to the conventional Q-learning algorithm, each training experience $\bm{e}_\ell=(\ell,\bm{s}_\ell,P_\ell,\bm{s}_{\ell+1},r_{\ell+1})$ is generated based on the $\xi$-greedy policy: given the current state $\bm{s}_\ell$, the policy selects the greedy action $P_\ell = \argmax_{P \in \mathcal{A}_\ell} Q(\bm{s}_\ell,P; \Theta)$ with probability $(1-\xi)$ or randomly chooses an action $P_\ell \in \mathcal{A}_\ell$ with probability $\xi$. To balance the exploration-exploitation, $\xi$ is initialized as $\xi_{\text{max}}$ and decreases linearly at each episode until it reaches the lower bound $\xi_{\text{min}}$, which is then maintained until the end of training (line 22 of Alg.~\ref{alg_3}). 

After updating the replay buffer $\mathcal{D}$, a random batch of experiences $\{\Tilde{\bm{e}}_1,\dots,\Tilde{\bm{e}}_{N_{\text{b}}}\}$ is sampled from $\mathcal{D}$ to update the primary Q-network, using the \ac{SGD} (lines 19-21 of Alg. 1). In particular, for a given experience $\Tilde{\bm{e}}_j = (\ell,\bm{s}_\ell,P_\ell,\bm{s}_{\ell+1},r_{\ell+1})$, the corresponding loss is defined as follows: first define the target Q-value of \ac{DDQN} as 
\begin{align}  \label{target_net_calc}
Y(\Tilde{\bm{e}}_j) = r_{\ell+1} + \hat{Q}(\bm{s}_{\ell+1},\argmax_{P \in \mathcal{A}_{\ell+1}} Q(\bm{s}_{\ell+1},P;\Theta);\hat{\Theta}).
\end{align}
Based on the Q-learning update in \eqref{basic_q_learn}, the loss associated with $\Tilde{\bm{e}}_j$ is then given by the mean-squared error between the target Q-value in \eqref{target_net_calc} and the predicted Q-value $Q(\bm{s}_\ell,P_\ell;\Theta)$, i.e., 
\begin{align} \label{Q_MSE_def}
L_{\Theta}(\Tilde{\bm{e}}_j) = \left[ Y(\Tilde{\bm{e}}_j) - Q(\bm{s}_\ell,P_\ell;\Theta) \right]^2.
\end{align}
When $\ell=L$, we set $Q(\bm{s}_{L+1},P_{L+1};\Theta) =\hat{Q}(\bm{s}_{L+1},P_{L+1};\hat{\Theta}) = 0$. With the loss given by \eqref{Q_MSE_def} and \eqref{target_net_calc}, the primary network parameter is then updated using the \ac{SGD}, as
\begin{align} \label{sgd_loss_sample}
\Theta  \leftarrow  \Theta - \nu \cdot \nabla_{\Theta} \left(\sum_{j=1}^{N_{\text{b}}} L_{\Theta} (\Tilde{\bm{e}}_j) \right).  
\end{align}

The pseudo-code for DDQN training is summarized in Algorithm 1. Given a trained primary Q-network $Q(\bm{s},P;\Theta^*)$, the causal power allocation is then given by  
\begin{align} \label{optimal_allocate_power_causal}
P_\ell = \argmax_{P \in \mathcal{A}_\ell} \ Q(\bm{s}_\ell,P;\Theta^*), \ \ell\in[L].
\end{align}

\begin{algorithm} 
\caption{\ac{DDQN} training for causal power allocation.} \label{alg_3}
\begin{algorithmic} [1]
\STATE \textbf{Channel parameters}: total power $P_0$, number of coherent blocks $L$, legitimate receiver $\text{SNR}_H$, warden $\text{SNR}_G$, covert constraint $\epsilon_\delta$
\STATE \textbf{Algorithm hyperparameters}: replay buffer size $N_{\text{buff}}$, number of $L$-state transitions $N_{\text{tr}}$, number of episodes per target Q-network update $N_{\text{ep}}$, batch size of experiences $N_{\text{b}}$, $\epsilon$-greedy parameters $(\xi_{\text{min}},\xi_{\text{max}},\Delta_{\xi})$, learning rate $\gamma$ 
\STATE Initialize the replay experience buffer $\mathcal{D}$, $\xi = \xi_{\text{max}}$
\STATE Randomly initialize network parameters $\Theta$ and $\hat{\Theta}$
\FOR{episode $E=0,1,2,\dots$}
    \FOR{state transition $n=1,\dots,N_{\text{tr}}$} 
        \STATE Randomly generate $\{h_\ell\}_{\ell=1}^{L}$ and $\{g_\ell\}_{\ell=1}^{L}$, let $P^{(1)}=P_0$, $\Gamma^{(1)}=0$ and $\epsilon^{(0)}=\epsilon_\delta$
        \FOR{$\ell = 1,\dots,L$}
            \STATE Generate the action $P_\ell$ based on $\xi$-greedy policy
            \IF{$\ell=L$}
                \STATE $P^{(L+1)}= \epsilon^{(L+1)}=\Gamma^{(L+1)}=0$ and $\bm{s}_{L+1}=\bm{0}$
            \ELSE
                \STATE Compute $P^{(\ell+1)}$, $\epsilon^{(\ell+1)}$, $\Gamma^{(\ell+1)}$ and $r_{\ell+1}$ based on \eqref{power_budget_property}, \eqref{covert_margin_property}, \eqref{cca_budget_property} and \eqref{reward_func_define}, respectively.
                \STATE If $|\mathcal{D}| \ge N_{\text{buff}}$, pop out the oldest experience
                \STATE Add experience $\bm{e}_\ell=(\ell,\bm{s}_\ell,P_\ell,\bm{s}_{\ell+1},r_{\ell+1})$ to $\mathcal{D}$
            \ENDIF
        \ENDFOR
    \ENDFOR
    \STATE Randomly select $N_{\text{b}}$ experiences $\{\Tilde{\bm{e}}_j\}_{j=1}^{N_{\text{b}}}$ from $\mathcal{D}$
    \STATE Compute the losses $\{L_{\Theta}(\Tilde{\bm{e}}_j)\}_{j=1}^{N_{\text{b}}}$ using \eqref{Q_MSE_def} and \eqref{target_net_calc}
    \STATE Update $\Theta$ by \ac{SGD} using \eqref{sgd_loss_sample}
    \STATE $\xi=\max\{\xi_{\text{min}},\xi_{\text{max}}-\Delta_{\xi} \cdot E\}$
    \STATE $\hat{\Theta} \leftarrow \Theta$ if $E = 0 \ \text{mod} \ N_{\text{ep}}$
\ENDFOR
\RETURN primary Q-network parameter $\Theta^*$
\end{algorithmic}
\end{algorithm}

\subsection{Rate Allocation Under Causal CSI} \label{c_bit_allo}
Unlike the causal power allocation scheme in Sec.~\ref{power_c_allo}, the rate allocation under causal \ac{CSI} cannot be formulated as an \ac{MDP} due to the constraint in \eqref{formu_bit_object_2}. Specifically, for power allocation, by defining the state variable $P^{(\ell)}$ in \eqref{power_budget_property}, the total power constraint in \eqref{formu_nc_nonconvex_AP} can be written as $P_\ell \leq P^{(\ell)}$, i.e., the current action is constrained by the current state. On the other hand, for rate allocation, the action variable is $R_\ell$ and if we similarly define the state variable as 
\begin{align} \label{rate_budget_property}
R^{(1)}=R_0, \  R^{(\ell+1)}=R^{(\ell)}-R_\ell, \ \ell \in [L-1],
\end{align}
then the total rate constraint becomes $\sum_{j=\ell}^{L} R_\ell \geq R^{(\ell)}$, i.e., the future actions are constrained by the current state. Hence the rate allocation is non-Markovian and therefore cannot be formulated as an \ac{MDP}.

Here we employ a trained power allocation network $Q(\bm{s},P;\Theta^*)$ in \eqref{optimal_allocate_power_causal} to approximately solve the causal rate allocation problem. Note that $R_\ell \triangleq \log\left(1+h_\ell P_\ell\right)$, hence we denote $\mathcal{R}(x)\triangleq\log(1+x)$ and $\mathcal{R}^{-1}(x)\triangleq 2^x-1$. We have
\begin{align} \label{lower_power_derive_1}
R^{(\ell)} \leq \sum_{i=\ell}^{L} & R_\ell = \sum_{i=\ell}^{L} \mathcal{R}(h_i P_i) \notag \\
&\leq (L-\ell+1) \cdot \mathcal{R}\left( \frac{\sum_{i=\ell}^{L} h_i P_i}{L-\ell+1} \right),
\end{align}
where the first inequality is due to \eqref{formu_bit_object_2}, and the second inequality follows from Jensen's inequality. Hence
\begin{align} \label{lower_power_derive_2}
\sum_{i=\ell}^{L} h_i P_i \geq (L-\ell+1) \cdot \mathcal{R}^{-1}\left( \frac{R^{(\ell)}}{L-\ell+1}\right).
\end{align}
Denote $\Bar{h} \triangleq \mathbb{E}(h)$ as the expected legitimate channel SNR. Then we approximate the left-hand side of \eqref{lower_power_derive_2} by $\sum_{i=\ell}^{L} h_i P_i \approx P^{(\ell)} \cdot \Bar{h}$, where $P^{(\ell)}$ is defined in \eqref{power_budget_property}. Therefore we can convert $R^{(\ell)}$ to $P^{(\ell)}$ as 
\begin{align} \label{lower_power_given_rate}
P^{(\ell)}(R^{(\ell)}) = \frac{L-\ell+1}{\Bar{h}} \cdot \mathcal{R}^{-1}\left(\frac{R^{(\ell)}}{L-\ell+1}\right).
\end{align}
In particular, we set $P_0 = P^{(1)}(R^{(1)}) = \frac{L}{\Bar{h}} \cdot \mathcal{R}^{-1}(\frac{R_0}{L})$ and train a primary Q-network $Q(\bm{s},P;\Theta^*)$ using Algorithm 1. Then we use the trained network to compute the causal rate allocations $R_\ell$ as follows. Starting from $R^{(1)}=R_0$, $\Gamma^{(1)}=0$ and $\epsilon^{(1)}=\epsilon_\delta$, for each block $\ell \in [L-1]$, given $(R^{(\ell)},\epsilon^{(\ell)},\Gamma^{(\ell)},h_\ell,g_\ell)$, we first compute $P^{(\ell)}(R^{(\ell)})$ in \eqref{lower_power_given_rate}, and the feasible set of power allocation $\mathcal{A}_\ell$ in \eqref{feasi_set_power_causal}, and select the power allocation $P_\ell$ using the trained network $Q(\bm{s},P;\Theta^*)$:
\begin{align} \label{rate_power_select}
P_\ell &= \argmax_{P \in \mathcal{A}_\ell} \ Q(\bm{s}_\ell, P; \Theta), \notag \\ 
\bm{s}_\ell &= (P^{(\ell)}(R^{(\ell)}),\epsilon^{(\ell)},\Gamma^{(\ell)},h_\ell,g_\ell).
\end{align}
Next, we compute $R_\ell(P_\ell)=\log\left(1+h_\ell P_\ell \right)$ and update
\begin{align} \label{rate_state_update}
R^{(\ell+1)} &= R^{(\ell)} - R_\ell(P_\ell), \ 
\Gamma^{(\ell+1)} = \Gamma^{(\ell)} + \log\left(\frac{1+h_\ell P_\ell}{1+g_\ell P_\ell}\right), \notag \\
\epsilon^{(\ell+1)} &= \epsilon^{(\ell)} - g_\ell P_\ell.
\end{align}
For the last block $L$, given $(R^{(L)},\epsilon^{(L)},\Gamma^{(L)},h_L,g_L)$ and $P^{(L)}(R^{(L)})=\frac{e^{R^{(L)}}-1}{h_L}$, we compute the feasible set $\mathcal{A}_L$ using \eqref{feasi_set_power_causal} and declare infeasibility if $\max_{P_L \in \mathcal{A}_L} \log\left(1+h_L P_L\right) < R^{(L)}$; otherwise we set $R_L=R^{(L)}$ and the above sequential rate allocation provides a feasible solution.

\section{Simulation results} \label{sec:num_result}

\subsection{Simulation Setup} \label{simu_setup}

\subsubsection{Quantized Channels} \label{quan_chan}
The legitimate receiver's and the warden's channels in \eqref{block_fad_basis} are generated as $H_\ell \sim \mathcal{CN}(0,\sigma_H^2)$ and $G_\ell \sim \mathcal{CN}(0,\sigma_G^2)$. The instantaneous SNRs $h_\ell=\frac{|H_\ell|^2}{\sigma_n^2}$ and $g_\ell=\frac{|G_\ell|^2}{\sigma_v^2}$ are quantized into 1024 levels. The average SNR of these two channels are defined as
\begin{align} \label{snr_two_chan_def}
\text{SNR}_H \triangleq 10 \log_{10} \left( \frac{\sigma_H^2}{\sigma_n^2} \right), \ \text{SNR}_G \triangleq 10 \log_{10} \left( \frac{\sigma_G^2}{\sigma_v^2} \right). 
\end{align}
We set the number of coherence blocks per codeword $L=10$, the number of symbols per coherence block $T=100$, and the noise powers $\sigma_n^2 = \sigma_v^2 = 1$. 

\subsubsection{Baseline Methods} \label{base_method}
We consider two baseline methods for solving the non-causal power allocation optimization problem in \eqref{formu_nc_nonconvex} if it is feasible: the first method, referred to as ``trivial'', is given by \eqref{suffi_p_allo_def}, and the second method, referred to as ``convex'', is given by the solution to the convex problem \eqref{formu_nc_nonconvex_obj}-\eqref{formu_nc_nonconvex_PBC} if this convex solution satisfies \eqref{formu_nc_nonconvex_CCA}, and otherwise it is given by \eqref{suffi_p_allo_def}. We also consider two baseline methods for solving the non-causal rate allocation optimization problem in \eqref{formu_nonconvex_bit_allo} if the infeasibility conditions in \eqref{power_feasi_condi} and \eqref{convex_rate_infea_condi} both do not hold: the first one, referred to as  ``trivial'', is given by \eqref{R_trivial_final_solu} if it satisfies \eqref{less_Noisy_Bit_NC}, otherwise it declares infeasibility; and the second one, referred to as ``convex'', is given by \eqref{convex_rate_solu_1} if this convex solution satisfies \eqref{less_Noisy_Bit_NC}, and otherwise it is given by the trivial baseline.

Also, two baseline methods are considered for solving the sequential power allocation problem under causal \ac{CSI}. The first method, referred to as ``average'', is proposed in \cite{chorti2015optimal}, which assigns the current power $P_\ell$ to maximize the total future rate by setting legitimate channel SNRs as the expected value, i.e., $h_j=\Bar{h}\triangleq\mathbb{E}(h)$, and the remaining power is evenly split, i.e., $P_j=\frac{P^{(\ell)}-P_\ell}{L-\ell}$, for $j=\ell+1,\dots,L$. That is, 
\begin{multline} \label{opt_power_allo_ave_general}
\text{$\hat{P}_{\ell} = \arg\max_{P \in \mathcal{A}_\ell} \ \bigg\{\log\left(1+Ph_{\ell}\right) +$} \\
\text{$\quad (L-\ell) \cdot \bigg[\log\left(1+\frac{P^{(\ell)} - P}{L - \ell} \cdot \Bar{h} \right)\bigg] \bigg\}.$}
\end{multline}
When $\ell=L$, $\hat{P}_L=\max \{P \in \mathcal{A}_L\}$. The other baseline, referred to as ``trivial'', is the causal version of the trivial non-causal baseline in \eqref{suffi_p_allo_def}: for each block $\ell$, $P_\ell=0$ if $h_\ell < g_\ell$ and $P_\ell=\frac{\epsilon_\delta}{|\mathcal{L} |g_\ell}$ otherwise, until reaching the total power constraint $P_0$. We also consider two baseline methods for the causal rate allocation problem. One termed ``average'' operates as follows \cite{chorti2015optimal}: given the remaining rate $R^{(\ell)}$ and the auxiliary variable $\Gamma^{(\ell)}$ for each block $\ell \in [L-1]$, it computes $P^{(\ell)}(R^{(\ell)})$ using \eqref{lower_power_given_rate} and $\hat{P}_\ell$ using \eqref{opt_power_allo_ave_general}, then it computes $R_\ell(\hat{P}_\ell)=\log\left(1+h_\ell \hat{P}_\ell\right)$ and updates $R^{(\ell+1)}$ and $\Gamma^{(\ell+1)}$ by \eqref{rate_state_update}. When $\ell=L$, it declares infeasibility if $\max_{P \in \mathcal{A}_L} \log\left(1+h_L P\right) < R^{(L)}$; otherwise it sets $R_L=R^{(L)}$ and obtains a feasible solution. The other baseline termed ``trivial'' operates as follows: for each block $\ell$, it assigns $R_\ell=0$ if $h_\ell < g_\ell$ and $R_\ell=\log\left(1+\frac{\epsilon_\delta h_\ell}{|\mathcal{L} |g_\ell}\right)$ otherwise, until reaching the total rate requirement $R_0$; if $R_0$ cannot be achieved, it declares infeasibility.

\subsection{Power Allocation Under Non-causal CSI} \label{noncausal_power_allo_simu}
\begin{figure*} [htbp]
     \centering
     \justifying
     \begin{subfigure}[b]{0.48\textwidth}
         \centering
         \includegraphics[width=7.8cm]{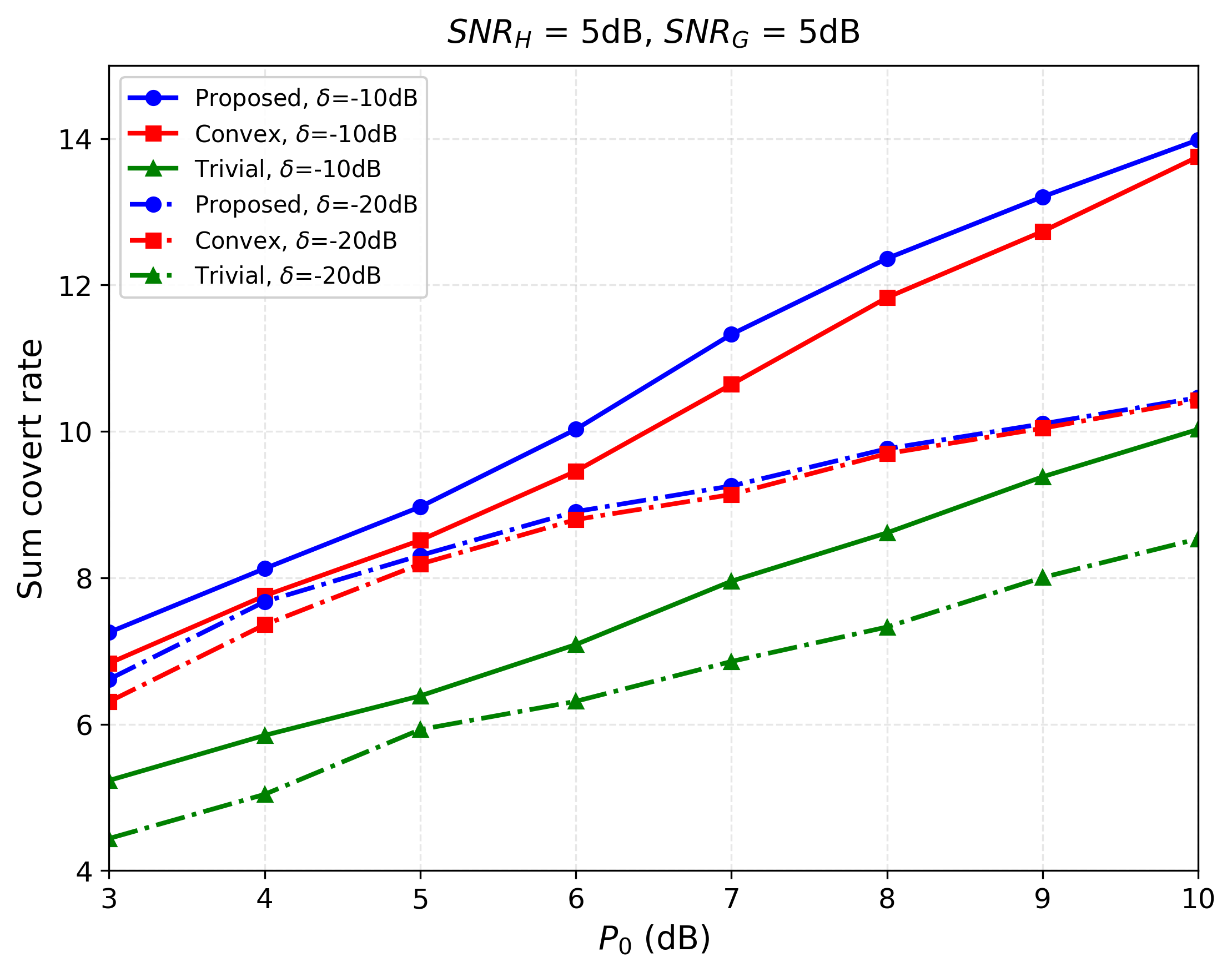}
            \caption{}
		\label{rate_diff_power_good}
     \end{subfigure}
     \begin{subfigure}[b]{0.48\textwidth}
        \centering
        \includegraphics[width=7.8cm]{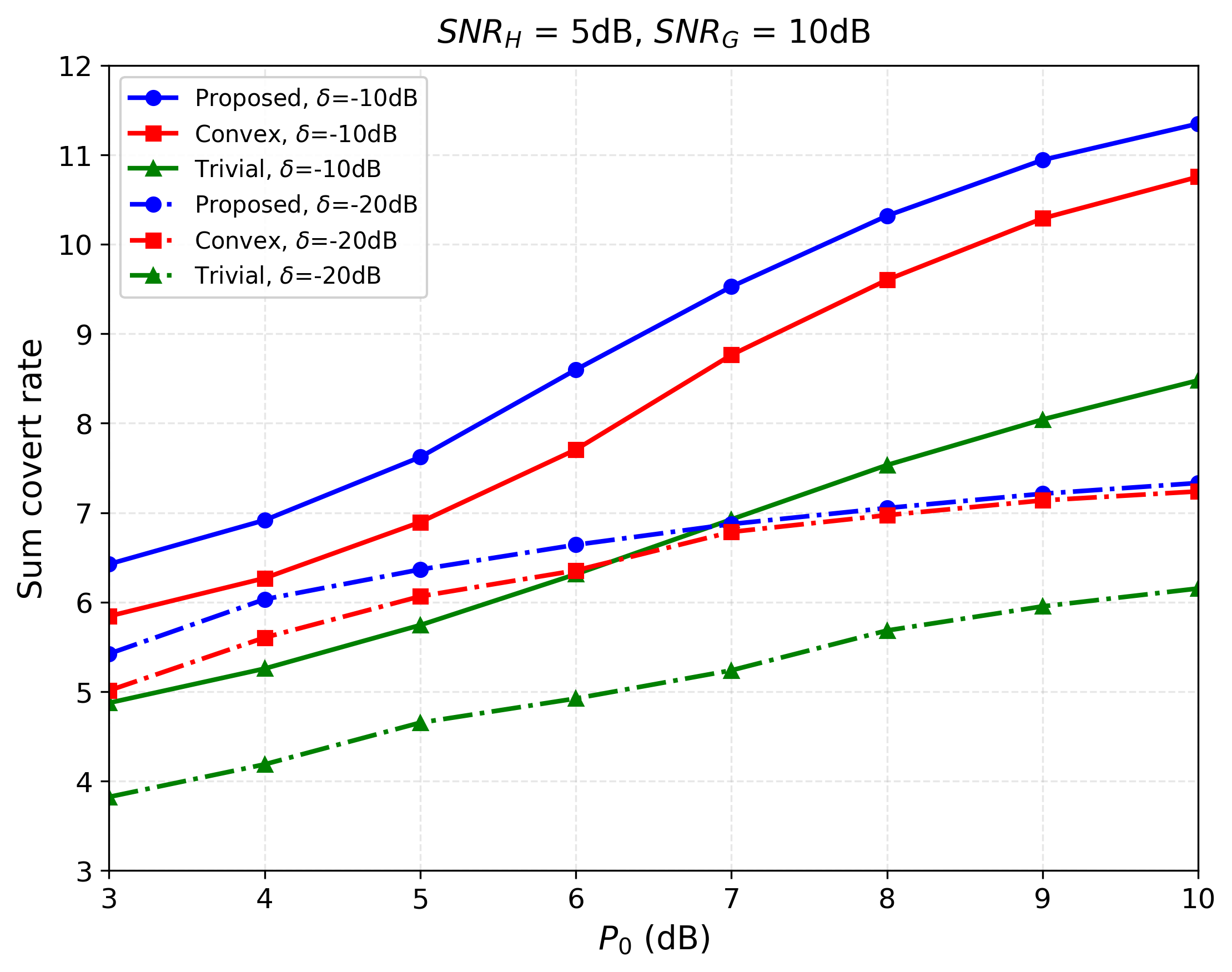}
            \caption{}
	    \label{rate_diff_power_bad}
    \end{subfigure}
        \caption{Sum covert rates under non-causal power allocation schemes.}
        \label{rate_diff_power}
\end{figure*}

We now illustrate the performance of our proposed algorithm in Sec.~\ref{power_allo_non_causal} for solving the power allocation problem in \eqref{formu_nc_nonconvex}. We set $\text{SNR}_H = 5\text{dB}, \text{SNR}_G = 5\text{dB}$ in \eqref{snr_two_chan_def}, $P_0=5\text{dB}$ in \eqref{formu_nc_nonconvex_AP}, $\delta=-10\text{dB}$ in \eqref{covertness_relaxed}, i.e., $\epsilon_\delta = 6.5\text{dB}$ in \eqref{formu_nc_nonconvex_PBC}, $C=20$ in \eqref{lam_origin_set}, $\alpha_1 = \alpha_2 = 10^{-6}$ in \eqref{proj_gd_define}, $N=10$ and $\gamma = 0.2$ in \eqref{lam_growth} and $\omega = 10^{-6}$ in \eqref{power_pga_stop_condi}. We first plot in Fig.~\ref{rate_diff_power} the average sum covert rate versus the total power constraint $P_0$ by the proposed method, the ``convex'' baseline method, and the ``trivial'' baseline method. The results indicate that the proposed scheme consistently outperforms the baseline schemes, while the convex baseline achieves a higher covert rate than the trivial baseline. Furthermore, the advantage of the proposed scheme becomes more pronounced when the warden has a better channel. Additionally, a smaller $\delta$ in \eqref{covertness_relaxed} results in a lower covert rate due to the stricter covertness constraint. 

\subsection{Rate Allocation Under Non-causal CSI} \label{noncausal_rate_allo_simu}
\begin{figure*} [htbp]
     \centering
     \justifying
     \begin{subfigure}[b]{0.48\textwidth}
         \centering
         \includegraphics[width=7.8cm]{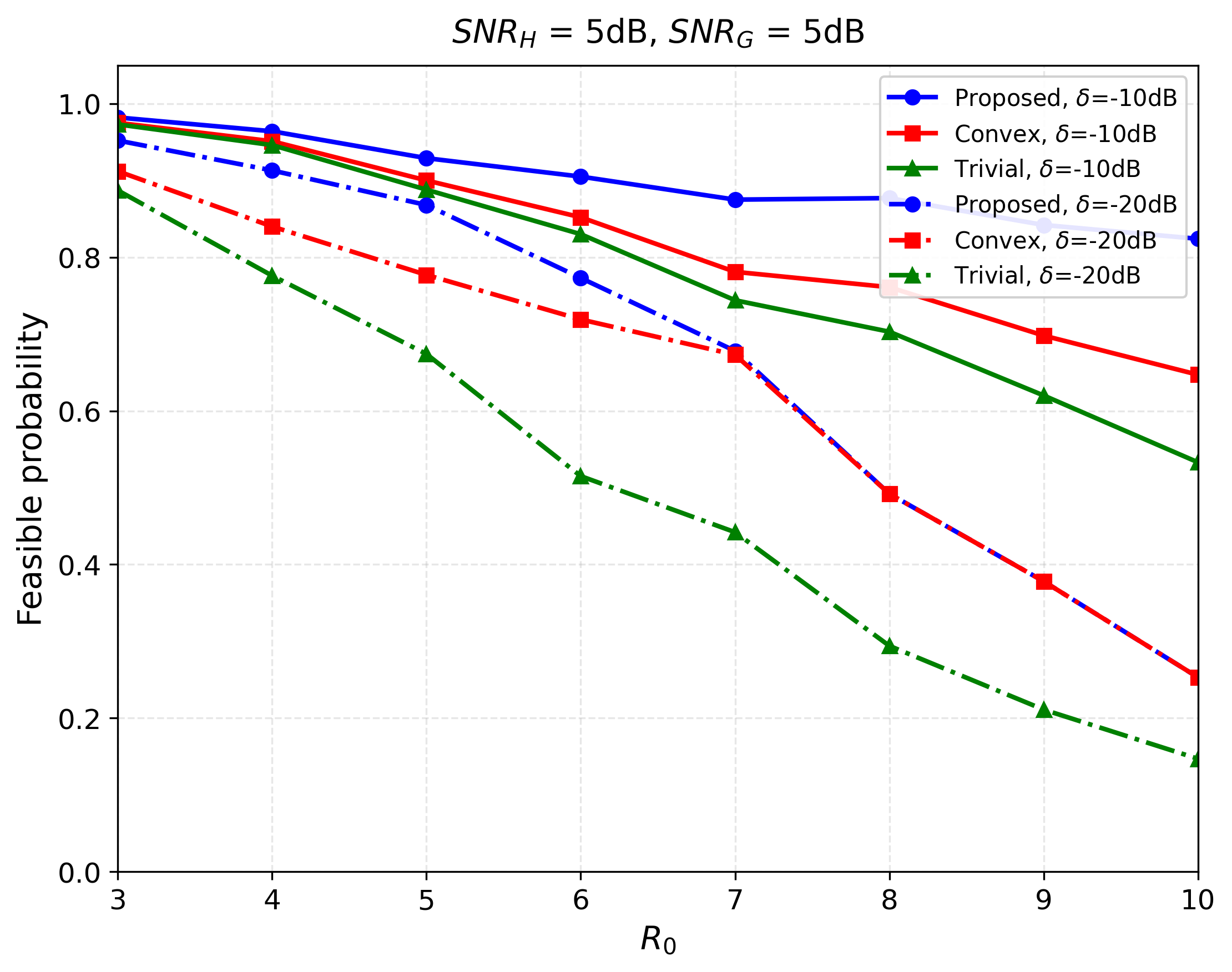}
            \caption{}
		\label{rate_allo_important_prob_good}
     \end{subfigure}
     \begin{subfigure}[b]{0.48\textwidth}
        \centering
        \includegraphics[width=7.8cm]{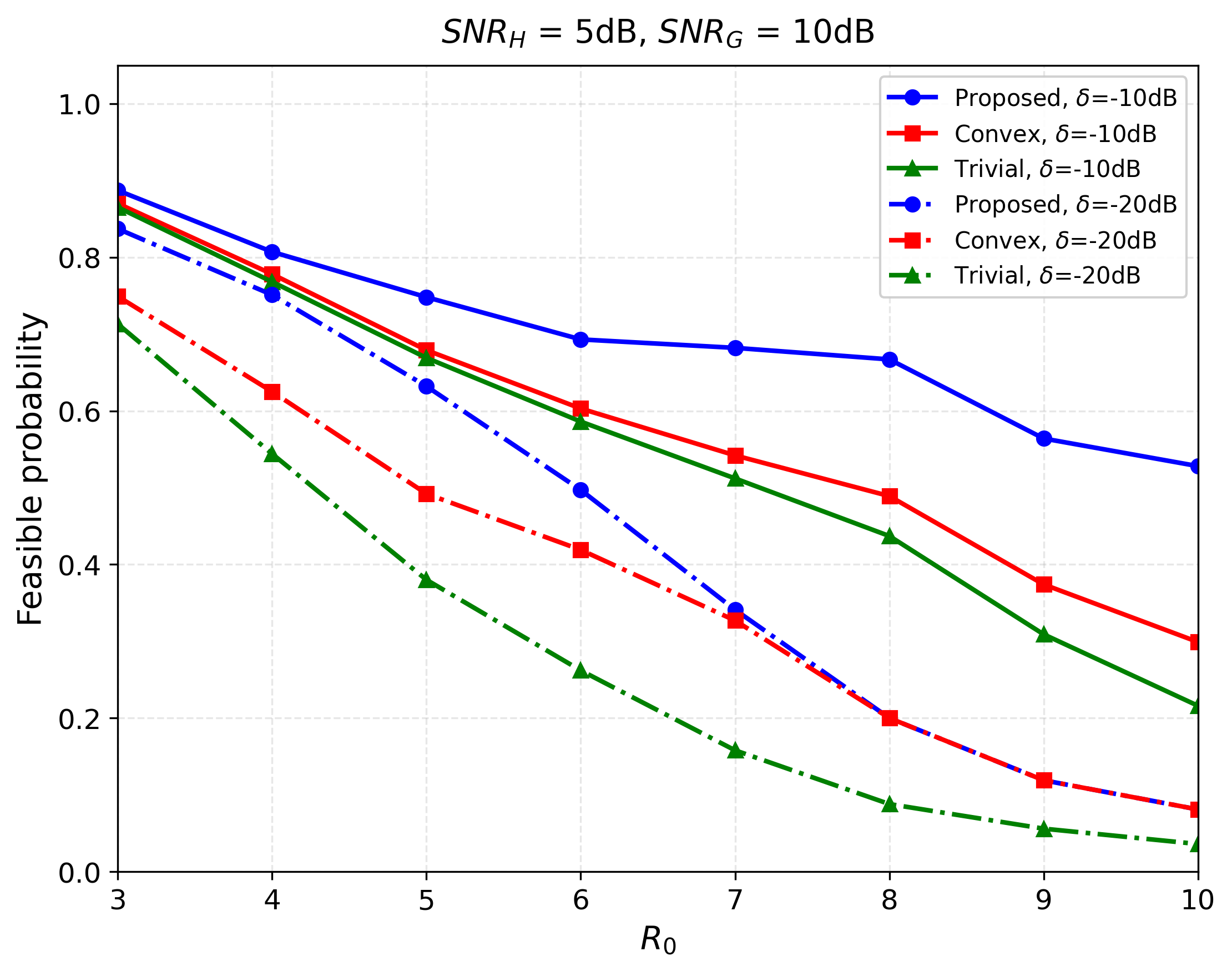}
            \caption{}
	    \label{rate_allo_important_prob_bad}
    \end{subfigure}
        \caption{Feasibility probabilities of non-causal rate allocation schemes.}
        \label{rate_allo_important_prob}
\end{figure*}

In this section, we illustrate the performance of the proposed algorithm in Sec.~\ref{bit_allo_non_causal} for solving the rate allocation problem in \eqref{formu_nonconvex_bit_allo}. We set $\text{SNR}_H = 5\text{dB}, \text{SNR}_G = 5\text{dB}$, $R_0=7$ in \eqref{formu_bit_object_2}, $\delta=-10\text{dB}$ in \eqref{covertness_relaxed}, i.e., $\epsilon_\delta = 6.5\text{dB}$ in \eqref{formu_bit_object_3}, $C=20$ in \eqref{rate_lam_origin_set}, $\alpha_1 = \alpha_2 = 10^{-6}$ in \eqref{rate_proj_gd_define}, $N=10$ and $\gamma = 0.2$ in \eqref{lam_growth} and $\omega = 10^{-6}$ in \eqref{rate_pgd_stop_condi}. We first plot in Fig.~\ref{rate_allo_important_prob} the feasibility probabilities of different schemes under two covertness constraints $\delta=-10$dB and $\delta=-20$dB. As before, we consider two SNR scenarios: (a) $\text{SNR}_H = \text{SNR}_G = 5$dB, and (b) $\text{SNR}_H = 5$dB, $\text{SNR}_G = 10$dB. It is observed that the feasibility probabilities become lower when the required sum rate $R_0$ increases. Comparing two sub-figures in Fig.~\ref{rate_allo_important_prob}, we find that when the warden channel is better than the legitimate channel, the feasibility probabilities decrease substantially for all three schemes. However, for both scenarios, the proposed scheme has a significantly higher feasibility probability than the ``convex'' and the ``trivial'' baseline methods. Furthermore, a smaller $\delta$ reduces the feasibility probabilities, since meeting the sum rate constraint in \eqref{formu_bit_object_2} becomes more challenging under a stricter covertness constraint.  

\begin{figure*} [htbp]
     \centering
     \justifying
     \begin{subfigure}[b]{0.48\textwidth}
         \centering
         \includegraphics[width=7.8cm]{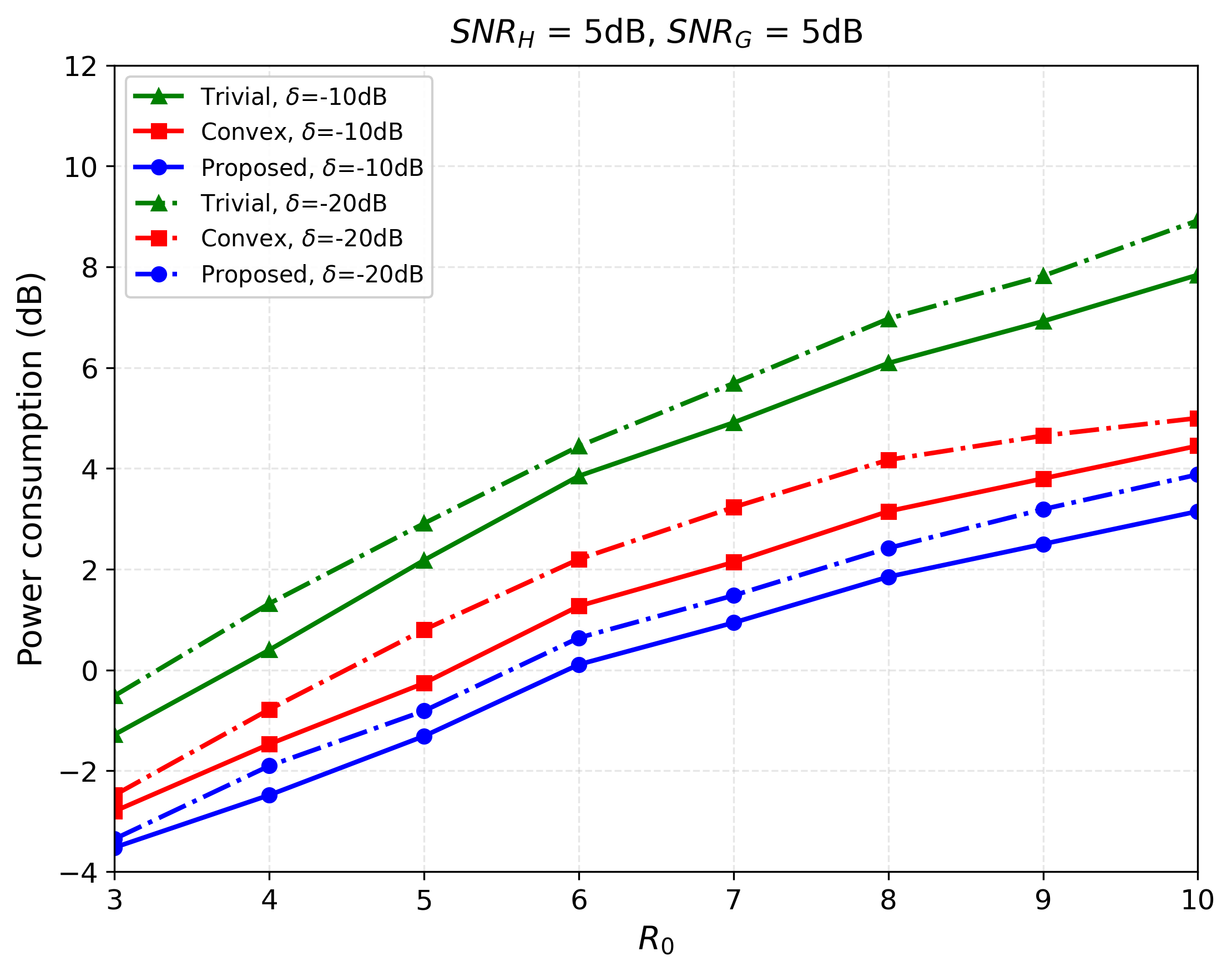}
            \caption{}
		\label{power_diff_rate_good}
     \end{subfigure}
     \begin{subfigure}[b]{0.48\textwidth}
        \centering
        \includegraphics[width=7.8cm]{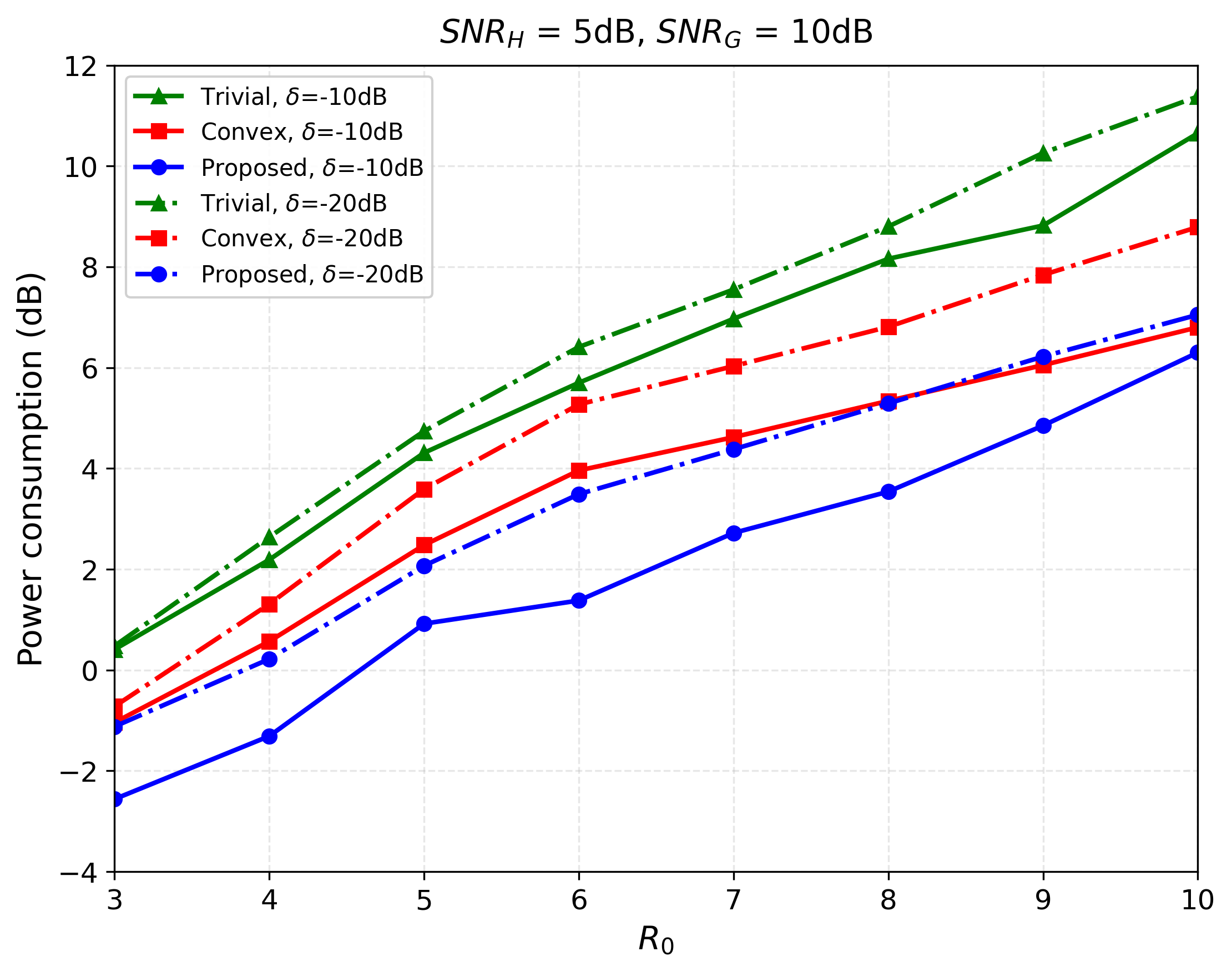}
            \caption{}
	    \label{power_diff_rate_bad}
    \end{subfigure}
        \caption{Power consumptions under non-causal rate allocation schemes when they are all feasible.}
        \label{power_diff_rate}
\end{figure*}

Next, in Fig.~\ref{power_diff_rate} we compare the average power consumptions when the proposed scheme and the two baselines all output feasible solutions. It is seen that the proposed scheme consistently achieves a lower power consumption than the two baselines, while the convex baseline consumes less power than the trivial baseline. Note that the actual sum rates achieved by the baseline schemes match the required rate $R_0$, since both the convex solution \eqref{convex_rate_solu_1} and the trivial solution \eqref{R_trivial_final_solu} satisfy $R_{\text{sum}}=R_0$; whereas the proposed scheme achieves a sum rate exceeding $R_0$, as the \ac{PGD} typically converges to a feasible rate allocation with  $R_{\text{sum}}>R_0$.  


\subsection{Power Allocation Under Causal CSI} \label{causal_power_allo_simu}

\begin{figure*}[htbp]
     \centering
     \justifying
     \begin{subfigure}[b]{0.48\textwidth}
         \centering
         \includegraphics[width=8.2cm]{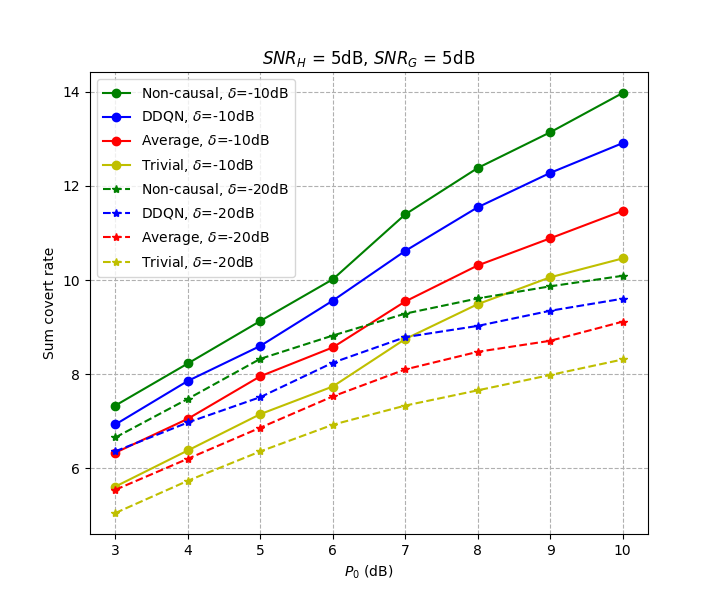}
            \caption{}
		\label{causal_power_good}
     \end{subfigure}
     \begin{subfigure}[b]{0.48\textwidth}
        \centering
        \includegraphics[width=8.2cm]{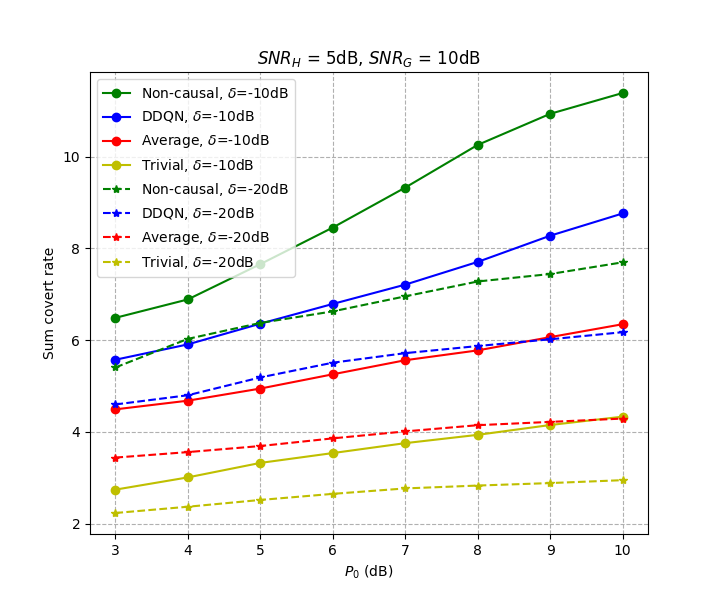}
            \caption{}
	    \label{causal_power_bad}
    \end{subfigure}
        \caption{Sum covert rates under causal power allocation schemes.}
        \label{causal_power_allo}
\end{figure*}

Now we illustrate the performance of our proposed \ac{DRL} algorithm in Sec.~\ref{power_c_allo} for solving the sequential power allocation problem under causal \ac{CSI}. To begin with, in Fig.~\ref{causal_power_allo} we plot the average sum covert rate versus the total power constraint $P_0$ of our proposed \ac{DDQN} causal power allocation and compare it with the ``average'' and the ``trivial'' baseline methods, as well as the proposed non-causal method in Sec.~\ref{power_allo_non_causal}. It is observed that the \ac{DDQN} power allocation scheme consistently outperforms the two causal baselines, and the performance gain becomes more pronounced when the warden has a better channel. On the other hand, compared with the non-causal power allocation, there is a covert rate loss due to the causality of the \ac{CSI}, which also increases when the warden's channel is better. Additionally, in both non-causal and causal cases, a smaller $\delta$ results in a lower covert rate corresponding to a stricter covertness constraint. 

\subsection{Rate Allocation Under Causal CSI} \label{causal_rate_allo_simu}
\begin{figure*}[htbp]
     \centering
     \justifying
     \begin{subfigure}[b]{0.48\textwidth}
         \centering
         \includegraphics[width=8.5cm]{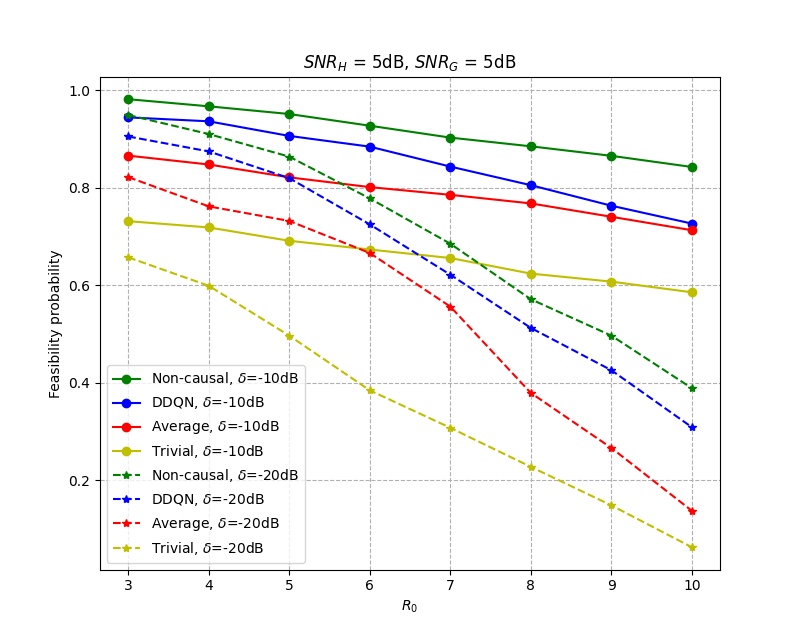}
            \caption{}
		\label{causal_rate_feasi_prop_good}
     \end{subfigure}
     \begin{subfigure}[b]{0.48\textwidth}
        \centering
        \includegraphics[width=8.5cm]{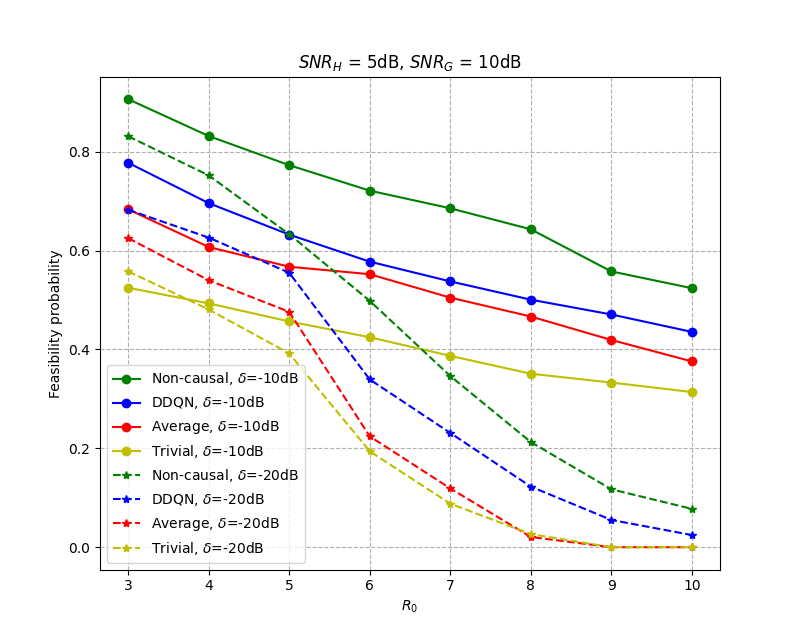}
            \caption{}
	    \label{causal_rate_feasi_prop_bad}
    \end{subfigure}
        \caption{Feasibility probabilities of causal rate allocation schemes.}
        \label{causal_rate_feasi_prop}
\end{figure*}

\begin{figure*} [htbp]
     \centering
     \justifying
     \begin{subfigure}[b]{0.48\textwidth}
         \centering
         \includegraphics[width=8.5cm]{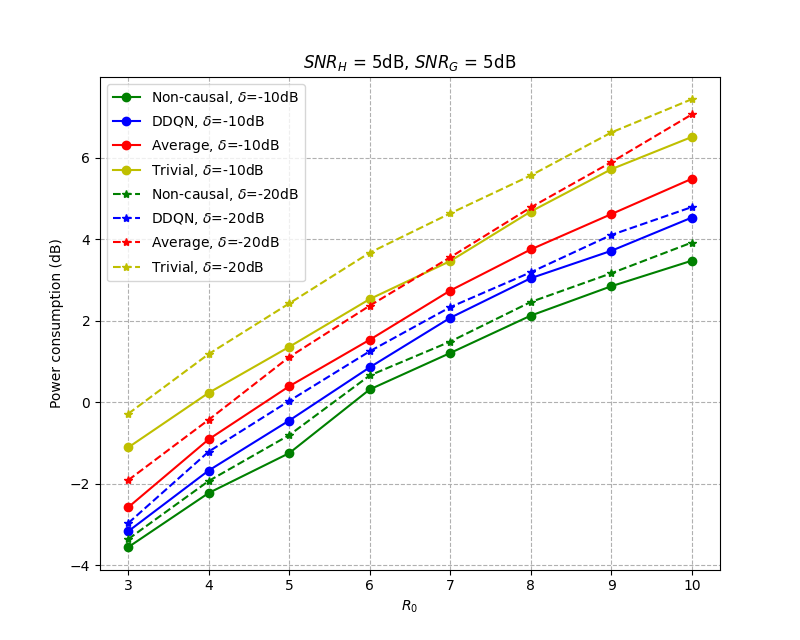}
            \caption{}
		\label{causal_rate_good}
     \end{subfigure}
     \begin{subfigure}[b]{0.48\textwidth}
        \centering
        \includegraphics[width=8.5cm]{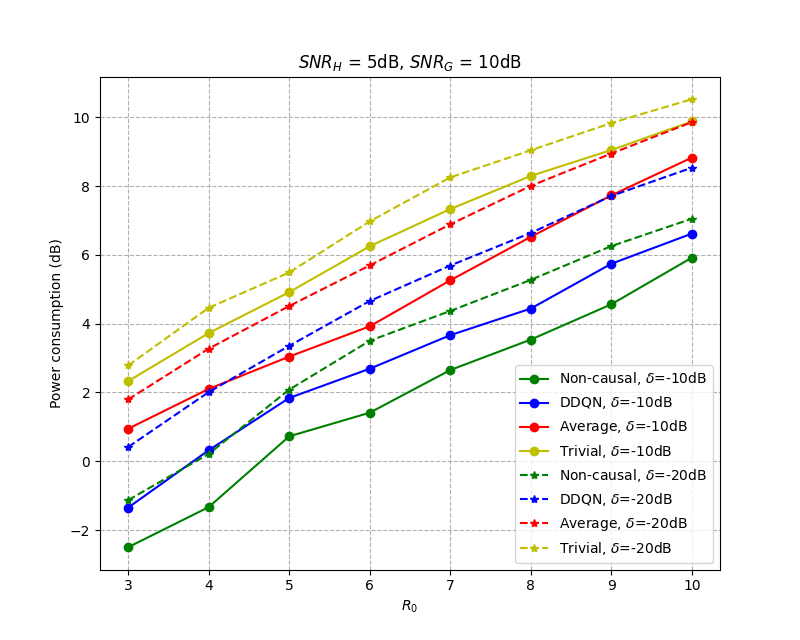}
            \caption{}
	    \label{causal_rate_bad}
    \end{subfigure}
        \caption{Power consumptions under causal rate allocation schemes when they are all feasible.}
        \label{causal_rate_allo}
\end{figure*}

Finally we demonstrate the performance of the proposed approximate \ac{DRL} algorithm in Sec.~\ref{c_bit_allo} for solving the sequential rate allocation problem under causal \ac{CSI}. Fig.~\ref{causal_rate_feasi_prop} shows the feasibility probabilities of the proposed \ac{DDQN}-based rate allocation method, the ``average'' and the ``trivial'' baseline methods, and the proposed non-causal method in Sec.~\ref{bit_allo_non_causal}. Fig.~\ref{causal_rate_allo} shows the average power consumption versus the total required rate $R_0$ for the same rate allocation schemes, under the condition that all schemes yield feasible solutions. It is observed from Fig.~\ref{causal_rate_allo} and Fig.~\ref{causal_rate_feasi_prop} that the \ac{DDQN} rate allocation scheme consistently outperforms two baseline methods, in terms of achieving both lower total power consumption and higher feasibility probability. On the other hand, the gaps to the corresponding non-causal rate allocation become larger when the warden's channel is better. Moreover, in both non-causal and causal cases, a smaller $\delta$ results in a higher power consumption and lower feasibility probability. 

\section{conclusions} \label{sec:conclu}
We have developed transmission strategies to achieve keyless positive-rate covert communication in block-fading channels. Based on the results developed for \acp{DMC}, we first formulate power allocation and rate allocation as non-convex optimization problems. When the \ac{CSI} of each fading block is known non-causally at the transmitter and legitimate receiver, we propose three-step methods to solve the corresponding power and rate optimization problems. For the case of causal \ac{CSI}, the corresponding power allocation problem is formulated as an \ac{MDP} and solved using a \ac{DDQN} method. Although the rate allocation problem under causal \ac{CSI} is not an \ac{MDP}, we solve it approximately using the \ac{DDQN} trained for power allocation. Extensive simulation results are provided to demonstrate the high performance of the proposed power and rate allocation algorithms to achieve keyless positive-rate covert communications, under both non-causal and causal \ac{CSI}.

\bibliographystyle{IEEEtran}
\bibliography{IEEEabrv,bibfile_2}

\end{document}